\documentclass[twoside,twocolumn,9pt]{article}
\usepackage{extsizes}
\usepackage[super,sort&compress,comma]{natbib}
\usepackage[version=3]{mhchem}
\usepackage[left=1.5cm, right=1.5cm, top=1.785cm, bottom=2.0cm]{geometry}
\usepackage{balance}
\usepackage{mathptmx}
\usepackage{sectsty}
\usepackage{graphicx}
\usepackage{lastpage}
\usepackage[format=plain,justification=justified,singlelinecheck=false,font={stretch=1.125,small,sf},labelfont=bf,labelsep=space]{caption}
\usepackage{float}
\usepackage{fancyhdr}
\usepackage{fnpos}
\usepackage[english]{babel}
\addto{\captionsenglish}{%
  
}
\usepackage{array}
\usepackage{droidsans}
\usepackage{charter}
\usepackage[T1]{fontenc}
\usepackage[usenames,dvipsnames]{xcolor}
\usepackage{setspace}
\usepackage[compact]{titlesec}
\usepackage{hyperref}
\usepackage{amsmath}

\usepackage{physics}
\usepackage{wasysym}
\usepackage{xcolor}
\usepackage{amsmath}
\usepackage{amssymb}
\usepackage{comment}
\usepackage{bm}

\definecolor{cream}{RGB}{222,217,201}

\begin{document}

\pagestyle{fancy}
\thispagestyle{plain}
\fancypagestyle{plain}{
    \renewcommand{\headrulewidth}{0pt}
}

\makeFNbottom
\makeatletter
\renewcommand\LARGE{\@setfontsize\LARGE{15pt}{17}}
\renewcommand\Large{\@setfontsize\Large{12pt}{14}}
\renewcommand\large{\@setfontsize\large{10pt}{12}}
\renewcommand\footnotesize{\@setfontsize\footnotesize{7pt}{10}}
\makeatother

\renewcommand{\thefootnote}{\fnsymbol{footnote}}
\renewcommand\footnoterule{\vspace*{1pt}%
    \color{cream}\hrule width 3.5in height 0.4pt \color{black}\vspace*{5pt}}
\setcounter{secnumdepth}{5}

\makeatletter
\renewcommand\@biblabel[1]{#1}
\renewcommand\@makefntext[1]%
{\noindent\makebox[0pt][r]{\@thefnmark\,}#1}
\makeatother
\renewcommand{\figurename}{\small{Fig.}~}
\sectionfont{\sffamily\Large}
\subsectionfont{\normalsize}
\subsubsectionfont{\bf}
\setstretch{1.125} 
\setlength{\skip\footins}{0.8cm}
\setlength{\footnotesep}{0.25cm}
\setlength{\jot}{10pt}
\titlespacing*{\section}{0pt}{4pt}{4pt}
\titlespacing*{\subsection}{0pt}{15pt}{1pt}
\fancyfoot{}
\fancyfoot[LO,RE]{\vspace{-7.1pt}\includegraphics[height=9pt]{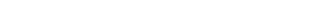}}
\fancyfoot[CO]{\vspace{-7.1pt}\hspace{13.2cm}\includegraphics{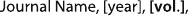}}
\fancyfoot[CE]{\vspace{-7.2pt}\hspace{-14.2cm}\includegraphics{head_foot/RF}}
\fancyfoot[RO]{\footnotesize{\sffamily{1--\pageref{LastPage} ~\textbar  \hspace{2pt}\thepage}}}
\fancyfoot[LE]{\footnotesize{\sffamily{\thepage~\textbar\hspace{3.45cm} 1--\pageref{LastPage}}}}
\fancyhead{}
\renewcommand{\headrulewidth}{0pt}
\renewcommand{\footrulewidth}{0pt}
\setlength{\arrayrulewidth}{1pt}
\setlength{\columnsep}{6.5mm}
\setlength\bibsep{1pt}

\makeatletter
\newlength{\figrulesep}
\setlength{\figrulesep}{0.5\textfloatsep}

\newcommand{\topfigrule}{\vspace*{-1pt}%
    \noindent{\color{cream}\rule[-\figrulesep]{\columnwidth}{1.5pt}} }

\newcommand{\botfigrule}{\vspace*{-2pt}%
    \noindent{\color{cream}\rule[\figrulesep]{\columnwidth}{1.5pt}} }

\newcommand{\dblfigrule}{\vspace*{-1pt}%
    \noindent{\color{cream}\rule[-\figrulesep]{\textwidth}{1.5pt}} }

\makeatother

\twocolumn[
    \begin{@twocolumnfalse}
        \vspace{1em}
        \sffamily
        \begin{tabular}{m{4.5cm} p{13.5cm} }
            \includegraphics{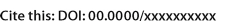}
            & \noindent\LARGE{\textbf{Flow states of two dimensional active gels driven by external shear}
            }\\
            \vspace{0.3cm} & \vspace{0.3cm}\\
            & \noindent\large{Wan Luo$^{\ast}$\textit{$^{ab}$}, Aparna Baskaran\textit{$^{c}$}, Robert A. Pelcovits\textit{$^{de}$}, and Thomas R. Powers\textit{$^{abde\dag}$}}\\
            \includegraphics{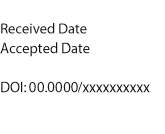} & \noindent\normalsize{Using a minimal hydrodynamic model, we theoretically and computationally study active gels in straight and annular two-dimensional channels subject to an externally imposed shear. The gels are isotropic in the absence of externally- or activity-driven shear, but have nematic order that increases with shear rate. 
             Using the finite element method, we determine the possible flow states for a range of activities and shear rates. Linear stability analysis of an unconfined gel in a straight channel shows that an externally imposed shear flow can stabilize an extensile fluid that would be unstable to spontaneous flow in the absence of the shear flow,
              and destabilize a contractile fluid that would be stable against spontaneous flow in the absence of shear flow. These results are in rough agreement with the stability boundaries between the base shear flow state and the nonlinear flow states that we find numerically for a confined active gel.  For extensile fluids, we find three kinds of nonlinear flow states in the range of parameters we study: unidirectional flows, oscillatory flows, and  dancing flows. To highlight the activity-driven spontaneous component of the nonlinear flows, we characterize these states by the average volumetric flow rate and the wall stress. For contractile fluids, we only find the linear shear flow and a nonlinear unidirectional flow in the range of parameters that we studied. For large magnitudes of the activity, the unidirectional contractile flow develops a boundary layer. Our analysis of annular channels shows how curvature of the streamlines in the base flow affects the transitions among flow states.
            }\\
        \end{tabular}
    \end{@twocolumnfalse} \vspace{0.6cm}
]

\renewcommand*\rmdefault{bch}\normalfont\upshape
\rmfamily
\section*{}
\vspace{-1cm}

\footnotetext{\textit{$^{\ast}$~Email: Wan\_Luo@brown.edu}}
\footnotetext{\textit{$^{\dag}$~Email: Thomas\_Powers@brown.edu}}
\footnotetext{\textit{$^{a}$~School of Engineering, Brown University, Providence, RI 02912, USA.}}
\footnotetext{\textit{$^{b}$~Center for Fluid Mechanics, Brown University, Providence, RI 02912, USA.}}
\footnotetext{\textit{$^{c}$~Martin Fisher School of Physics, Brandeis University, Waltham, MA 02453 USA.}}
\footnotetext{\textit{$^{d}$~Department of Physics, Brown University, Providence, RI 02912, USA.}}
\footnotetext{\textit{$^{e}$~Brown Theoretical Physics Center, Brown University, Providence, RI 02912, USA.}}



\newcommand\trp[1]{#1}
\newcommand\rap[1]{#1}
\newcommand\wl[1]{#1}
 \newcommand{\trpb}[1]{{\textcolor{blue}{#1}}}
\newcommand{\abeds}[1]{{\textcolor{black}{#1}}}



\section{Introduction}
\label{sec:introduction}


The defining property of an active fluid is that energy is added to the system at the  small length scales of the particles that make up the fluid, instead of at the large length scales of the bounding walls or inlets of the system.\cite{marchetti2013hydrodynamics}
Commonly studied examples include cytoplasm~\cite{GoldsteinTuvalvandeMeent2008} or its reconstituted components,\cite{NedelecSurreyMaggsLeibler1997,sanchez2012spontaneous,Alvarado2017} collections of swimming microorganisms,\cite{RiedelKruseHoward2005,Koch2011,Saintillan2013} and model two-dimensional  layers of cells.\cite{Duclos_etal2016}
The interplay of the energy injected at small scales and the interactions among the constituent particles lead to nonequilibrium collective behavior, including spontaneous coherent flows,\cite{woodhouse2012spontaneous,lushi2014fluid,wu2017transition} sustained oscillations,\cite{marchetti2013hydrodynamics,samui2021flow} active turbulence,\cite{wensink2012meso,dombrowski2004self,dunkel2013fluid} and two-dimensional\cite{sanchez2012spontaneous} or three-dimensional~\cite{simha2002hydrodynamic,vcopar2019topology}
topological defects in active liquid crystalline fluids.  These phenomena suggest that active fluids may be used for novel microfluidics applications, including fluids that pump themselves or mix themselves. Since these applications require a degree of control over active fluids, recent investigations have studied how confinement of active fluids affects flows and the formation of defects.\cite{Arajo2023,norton2018insensitivity,opathalage2019self, samui2021flow,hardouin2019reconfigurable}
In this paper, we build on 
these investigations by studying the flow states of an active gel in a channel with moving boundaries to see how an imposed shear affects the possible flow states and the transitions among them.  

By `active gel' we mean a model liquid crystal which tends to the isotropic phase away from boundaries with strong anchoring conditions and in the absence of shear flow. The motionless, isotropic state of an unbounded two-dimensional active gel is unstable to spontaneous flow and nematic ordering above a critical activity.\cite{Soni2018,Santhosh_etal2020} Recent numerical calculations have identified the spontaneous flow states in straight three-dimensional~\cite{VargheseBaskaranHaganBaskaran2020,chandrakar2020confinement} and two-dimensional channels~\cite{VargheseBaskaranHaganBaskaran2020,samui2021flow,chandragiri2020flow} with stationary walls. In a two dimensional channel with no-torque anchoring conditions at the  walls, the critical activity for spontaneous flow increases as the channel width decreases.\cite{VargheseBaskaranHaganBaskaran2020} Thus, confinement is stabilizing, as has been found in other related situations.\cite{wioland2016directed} For a given value of the activity parameter, new flow states emerge as the channel width increases, with the flow progressing through unidirectional, undulating (also known as `oscillatory',\cite{samui2021flow}) and dancing flow states.\cite{shendruk2017dancing,VargheseBaskaranHaganBaskaran2020} A similar sequence of flow states is found for fixed channel width and increasing activity.\cite{VargheseBaskaranHaganBaskaran2020}

Our work is  motivated by the experimental observation that imposed shear can prevent~\cite{chandrakar2020confinement} the spontaneous instability of a solution~\cite{wu2017transition} of microtubule bundles and kinesin motors in the presence of the molecular fuel ATP.  Instead of a motionless state, our base state is the state of simple shear in which the flow field is given by the solution to the Stokes equation for our straight or annular channel geometry. Working at fixed channel width, we find that increasing the activity leads to a sequence of flow states which are reminiscent of the ones seen in the case of no external shear, but with some important new elements. For example, the imposed shear rate can be stabilizing in the same sense that confinement is stabilizing: for an extensile active gel, we find that the critical activity for the imposed simple shear flow to develop a spontaneous flow component increases with the imposed shear rate. A similar result was 
established using 
linear stability analysis of a polar system by Muhuri, Rao, and Ramaswamy.\cite{MuhuriRaoRamaswamy2007} Here we give a more systematic treatment of this problem for the apolar case, revealing that the imposed shear also leads to oscillatory behavior in the unstable modes. For a contractile active gel, we find that shear is \textit{destabilizing}. Earlier work has also examined the rheology of active nematics and gels, showing that polar active particles have a nonmonotonic stress-strain relation at high activity,\cite{GiomiLiverpoolMarchetti2010} and illuminating the nature of shear banding in apolar active gels.\cite{Fielding2011}  Our work extends these investigations to the case of an annular channel, illustrating the role of the curvature of the streamlines of the base flow. 

Our paper begins with 
a minimal hydrodynamic model for active gels. 
We then study the linear stability of an active gel in a straight channel subject to a uniform shear flow imposed by a moving plate. In the stable region, the linear rheology, orientational order, and the shear stress exerted by the active fluids on the moving boundary are analytically calculated for the state of uniform shear. Then we turn to the other flow states 
using the finite element method to characterize the flow transitions 
for the extensile and contractile fluids. Next, we turn to an annular channel and carry out similar analytical and numerical studies to assess the effects of the curvature of the boundaries.


\section{Minimal hydrodynamic model}
\label{model}
We use a well-\trp{studied} 
continuum hydrodynamic model
for nematic liquid crystals~\cite{olmsted1992isotropic,toth2002hydrodynamics} to describe apolar microtubules, adding a term corresponding to non-equilibrium active forces as was done in the "minimal" model \trp{used} 
by Varghese et al.\cite{VargheseBaskaranHaganBaskaran2020}
In two dimensions, the orientational order of 
apolar active matter is described by a traceless, symmetric tensor---the 
tensor order parameter that is used in the theory of nematic liquid crystals---$Q_{ij}=S( 2 n_i n_j-\delta_{ij})$, with $i,j=x,y$.~\cite{deGennesProst} 
The unit vector $\mathbf{n}(\mathbf{x})$ is the director at position $\mathbf{x}$ and the scalar order parameter $S$ represents the degree of alignment.
The equilibrium state of the microtubule  bundles is governed by a Landau-Ginzburg free energy density,
\begin{equation}
\begin{aligned}
\mathcal{F}&=\frac{K}{2}\partial_i Q_{jk}\partial_i Q_{jk}+\frac{A}{2}Q_{ij}Q_{ij}+\frac{C}{4}\left(Q_{ij}Q_{ij}\right)^2,
\end{aligned}
\label{LG}
\end{equation}
where repeated indices are summed over. 
The single Frank elastic constant $K$ penalizes gradients of $Q_{ij}$.  Since we focus on a low concentration isotropic phase, 
$A$ will be positive to guarantee that the minimizing state is disordered.
In two dimensions there is no term cubic in  $Q_{ij}$, and the isotropic-nematic transition is continuous. In the isotropic phase we consider in this paper, the term proportional to $C$ can be neglected, as was done in
previous studies of two-dimensional and three-dimensional channel flow.\cite{VargheseBaskaranHaganBaskaran2020,chandrakar2020confinement}


A minimal hydrodynamic model for incompressible flow in two dimensions is given by\cite{VargheseBaskaranHaganBaskaran2020}
\begin{eqnarray}
0&=&\boldmath{\nabla}\cdot\mathbf{v} \label{incompress}\\
0&=&-\boldsymbol{\nabla}p+\eta\nabla^2\mathbf{v}-a\nabla\cdot\textsf{Q}\label{veqn}\\
0&=&-\nu\big(\partial_t\textsf{Q}+ \textbf{v} \cdot \nabla \textsf{Q}+\textsf{Q} \cdot \Omega-\Omega \cdot \textsf{Q}\bigr)-A\textsf{Q}\nonumber+K\nabla^2\mathsf{Q} \\
&+&2 \lambda\nu \mathsf{E},\label{Qeqn}
\end{eqnarray}
where 
$\eta$ is the shear viscosity, $\nu$ is the rotational viscosity, $p$ is pressure, $(\mathbf{v}\cdot\grad{\mathsf{Q}})_{ij}=v_k \partial_k Q_{ij}$,  $\mathsf{E}=(\nabla \mathbf{v}+(\nabla \mathbf{v})^{\rm{T}})/2$ is the strain rate tensor, $\mathsf{\Omega}=(\nabla \mathbf{v}-(\nabla \mathbf{v})^{\rm{T}})/2$ 
[i.e. $\Omega_{ij}=(\partial_j v_i-\partial_i v_j)/2$]
is the vorticity tensor, and $a$ is the strength of the activity. A positive value of $a$ corresponds to extensile particles, and a negative value of $a$ corresponds to contractile particles. The shape parameter $\lambda$ is positive for prolate particles and negative for oblate particles; $\lambda=1$ corresponds to needle-like particles. 
Note that in three dimensions there will be additional nonlinear terms proportional to $\lambda$ appearing in eqn~(\ref{Qeqn}).

We disregard inertial effects because the Reynolds number of the typical active flows we study is small. In this minimal hydrodynamic model, passive backflow effects are neglected and the order parameter field $\mathsf{Q}$ only affects the flow through the active stress $-a\mathsf{Q}$. 
\begin{figure}[t]
\centering
\includegraphics[width=3.5in]{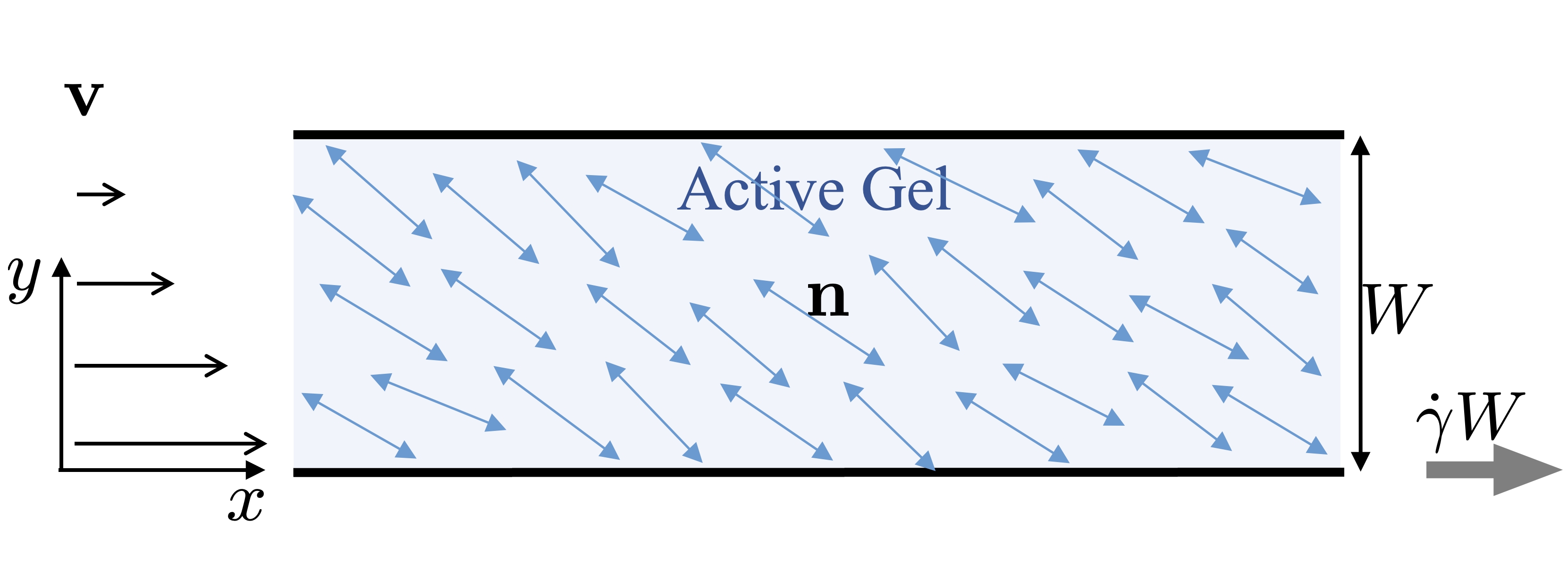}
\caption{The base state 
for the flow \trp{field} and \trp{tensor order parameter} field $Q_{ij}$ of an active gel in a straight channel with the bottom wall moving at a fixed speed $\dot{\gamma}W$. The 
\trp{double-headed arrows} correspond to the director field $\mathbf{n}$ of the extensile apolar active bundles. 
\trp{The tensor order parameter field is uniform throughout the channel because the flow is uniform and because we impose Neumann boundary conditions on $Q_{ij}$.
}
}
\label{fig:eg}
\end{figure}
The active time scale which results from the competition between viscosity and activity is given by $\eta/|a|$.
From the dynamical equation for $\mathsf{Q}$, eqn~(\ref{Qeqn}), it is apparent that the relaxation time $\tau$ for distortions away from the equilibrium isotropic 
state is $\tau=\nu/A$. Likewise, $\sqrt{K/A}$ is a correlation length for the liquid crystalline order, which we write in nondimensional form as $\ell=\sqrt{K/A}/W$\rap{, where $W$ is the width of the straight or annular channel}.
The factor $\lambda\nu$ characterizes the flow birefringence of a passive ($a=0$) liquid crystal.\cite{DeGennes1969}
When weak shear $\dot{\gamma}\ll1/\tau$ is applied to a nematic liquid crystal in the isotropic state, the rods align such that $A\mathsf{Q}\approx2\lambda\nu\mathsf{E}$, which implies that the scalar order parameter is proportional to the shear rate: $S\propto\dot{\gamma}\tau$.

\section{Straight channel: start-up problem and linear stability analysis}
\label{stability}

\abeds{Let us begin by} 
review\abeds{ing} the linear stability analysis of an unbounded two-dimensional active gel \cite{hatwalne04}. 
An isotropic ($\mathsf{Q}=0$),  motionless ($\mathbf{v}=0$) gel is unstable to shear flow and nematic ordering when the effective shear viscosity ($\eta_\mathrm{eff}\equiv\eta-a \lambda \tau$) vanishes, which occurs for a critical activity $a_\mathrm{c}=\eta/(\lambda\tau)$.\cite{Soni2018,Santhosh_etal2020} 
The form of the effective shear viscosity shows that extensile particles tend to reduce the shear viscosity, whereas contractile particles tend to increase it. 
In the  unstable state of the unconfined geometry, the pattern of alignment of the bundles follows a sine wave, appearing like a bent filament, or like the nematic configuration of bend.\cite{deGennesProst}


Next, \abeds{let us} consider 
an active gel confined to an infinite straight channel of 
width $W$ and subject to a steady uniform shear flow $\mathbf{v}_0=\dot{\gamma}(W-y)\hat{\mathbf{x}}$ as shown in Fig. \ref{fig:eg}. We assume no-slip boundary conditions on the channel walls for the velocity field, and Neumann conditions, ($\partial_i Q_{jk}=0$) or ``zero-torque conditions"  for the order parameter field on the walls. 
Given the parallel planar channel walls and zero-torque  boundary conditions, the nematic order parameter is uniform and divergenceless for the imposed uniform shear flow. In our hydrodynamic model, activity only appears in eqn (\ref{veqn}), and thus, when activity is below the critical value for the instability, the order parameter field is unaffected by the activity. 

\begin{figure}[t]
\centering
\includegraphics[height=1.8in]{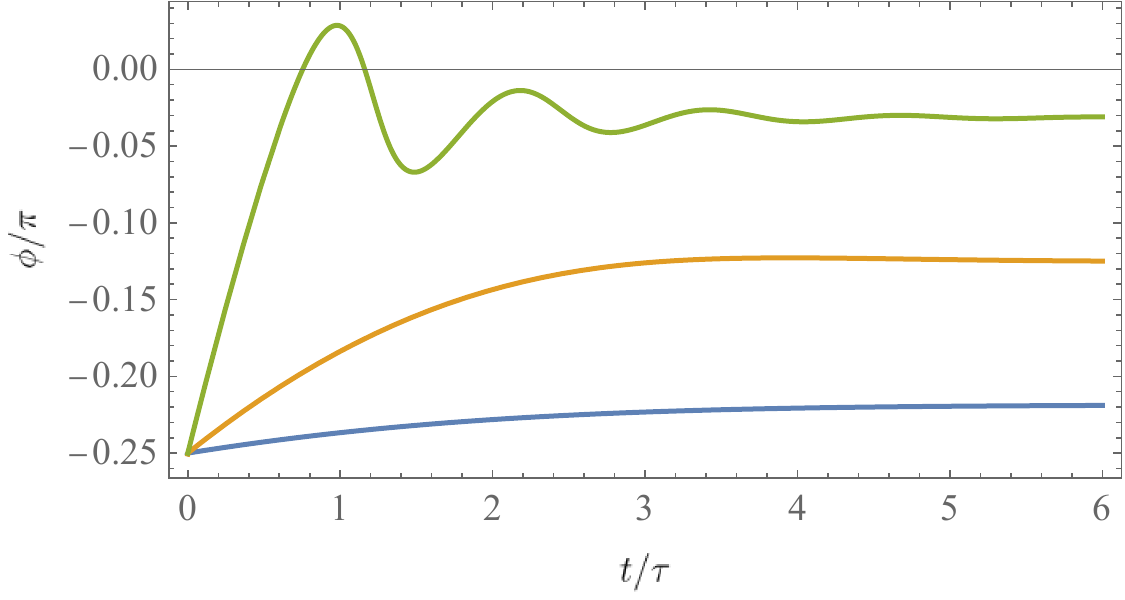} 
\caption{
Director angle $\phi$ 
as a function of time for various shear rates for the startup problem of the liquid crystal order parameter in the  case of steady simple shear. From top to bottom, 
the shear rates are $\dot{\gamma}\tau=5$ (green curve), $\dot{\gamma}\tau=1$ (gold curve), and $\dot{\gamma}\tau=0.2$ (blue curve).
}
\label{fig:oscQ}
\end{figure}
Before considering the stability of simple shear flow, we solve the startup problem, assuming an initially stationary isotropic gel with activity below the critical value (to be deduced below). Since the Reynolds number is assumed to be small, the flow immediately assumes its steady-state value $\mathbf{v}_0$. But the order parameter field attains its steady-state value only after a time comparable to the liquid crystal relaxation time $\tau$.~\cite{KriegerDiasPowers2015} Given the boundary conditions on the order parameter, we may assume that $\mathsf{Q}$ is uniform in space. Since $\mathsf{Q}$ is uniform, the divergence of the active stress vanishes and the flow remains simple shear as the order-parameter field evolves. The order parameter equations~(eqn (\ref{Qeqn})) reduce to 
\begin{eqnarray}
\partial_t Q_{xx}&=&-\frac{1}{\tau}Q_{xx}-\dot{\gamma} Q_{xy}\\
\partial_t Q_{xy}&=&\dot{\gamma} Q_{xx}-\frac{1}{\tau}Q_{xy}-\lambda\dot{\gamma}.
\end{eqnarray}
Assuming $\mathsf{Q}(t=0)=0$,  we find
\begin{eqnarray}
    Q_{xx}&=&Q^{(0)}_{xx}\left[1-\mathrm{e}^{-t/\tau}\cos\left(\dot{\gamma} t \right)\right]
+Q^{(0)}_{xy}\mathrm{e}^{-t/\tau}\sin\left(\dot{\gamma} t \right)\label{Qxxstartup}\\
Q_{xy}&=&Q^{(0)}_{xy}\left[1-\mathrm{e}^{-t/\tau}\cos\left(\dot{\gamma} t \right)\right]
-Q^{(0)}_{xx}\mathrm{e}^{-t/\tau}\sin\left(\dot{\gamma} t \right),\label{Qxystartup}
\end{eqnarray}
where the steady-state order parameter tensor $\mathsf{Q_0}$ is given by
\begin{eqnarray}
Q^{(0)}_{xx}&=&\frac{\lambda\dot{\gamma}^2\tau^2}{1+\dot{\gamma}^2\tau^2},\label{Qxx0}\\
Q^{(0)}_{xy}&=&-\frac{\lambda\dot{\gamma} \tau}{1+\dot{\gamma}^2\tau^2}.\label{Qxy0}
\end{eqnarray}
The order parameter rises to its steady state, with oscillations that become apparent when the shear rate is greater than the relaxation rate $1/\tau$. These oscillations are reminiscent of the oscillations observed~\cite{GuJamiesonWang1993} in the apparent viscosity during the startup flow of 8CB, a director-tumbling nematogen.\cite{larson1999} In simple shear, the director of a tumbling nematic makes a complete revolution, like a rod undergoing a Jeffery orbit in shear flow.\cite{larson1999}  In our case, as long as $\tau$ is finite, the directors oscillate about their final steady state. Fig.~\ref{fig:oscQ} shows the 
director angle $\phi=\arctan[Q_{xy}/(S+Q_{xx})]$ (measured counterclockwise from the $x$-axis) as a function of time. 

The steady-state scalar order parameter and the director angle 
are given by
\begin{eqnarray}
S&=&
\frac{\lambda\dot{\gamma} 
\tau}{\sqrt{1+\dot{\gamma}^2\tau^2}}\label{eq:S}\\ 
\phi 
&=&-\arctan \left(\frac{1}{\sqrt{1+\dot{\gamma}^2 \tau^2}+\dot{\gamma}\tau}\right ).\label{eq:phi} 
\end{eqnarray}
Equations~(\ref{eq:S}) and (\ref{eq:phi}) show that in steady state, the flow aligns the nematic director at a nonzero angle with the horizontal streamlines, with a degree of order that increases with increasing shear rate. 
At low shear rates, $\dot{\gamma}\tau \ll 1$, the bundles are oriented at an angle of 
$\phi=-\pi/4$ with the streamlines, and the order is weak ($S\ll1)$. At high shear rates, the bundles tend to align parallel to the streamlines, and $S\approx\lambda$. For needle-like particles, with $\lambda\approx 1$, the order is strong in the limit of high shear rate.
The shear stress on the moving plate in the stable region is 
\begin{equation}
\sigma_{\rm{W}}=-\eta \dot{\gamma}-a Q^{(0)}_{xy}=\dot{\gamma}\left(-\eta+\frac{a\lambda \tau}{1+\dot{\gamma}^2\tau^2}\right),\quad a<a_c. \label{wallshear}   
\end{equation}
From eqn~(\ref{wallshear}), it is easy to see the wall shear stress increases linearly with activity but the dependence on the imposed shear is not linear when the activity is below the critical value.


To analyze the 
stability 
of 
the base configuration 
with flow rate $\mathbf{v}_0$ \abeds{and the confinement $W$} 
, 
we consider 
a perturbation that is independent of $x$, \abeds{the channel axis}. \footnote{A more general assumption would be to suppose the perturbation depends on both $x$ and $y$, but here we forbid $x$-dependence to simplify the analysis. The more general analysis using pseudospectral methods will be reported elsewhere.}  
Thus, $\mathbf{v}=\mathbf{v}_0+ \mathbf{v}_1$ and $\mathsf{Q}=\mathsf{Q}_0+ \mathsf{Q}_1$ , with the perturbations
\begin{eqnarray}
  \mathbf{v}_1&=& 
  v_x \sin(n\pi y/W)\exp\left(\beta t\right) \hat{\mathbf x},\label{dimensionv}\\
  \mathsf{Q}_1&=&\begin{pmatrix} \mathcal{Q}_{xx} & \mathcal{Q}_{xy}\\ \mathcal{Q}_{xy} & -\mathcal{Q}_{xx}\end{pmatrix} \cos(n\pi y/W)\exp\left(\beta t\right),
\end{eqnarray}
where $v_x$, $\mathcal{Q}_{xx}$, and $\mathcal{Q}_{xy}$ are constants, $n$ is a nonzero positive integer, and $\beta$ is the growth rate of the perturbation.
With these assumptions, the $x$ component of the force equation eqn~(\ref{veqn}) implies 
\begin{equation}
{v}_x=\frac{a \mathcal{Q}_{xy} W}{n \pi \eta}.\label{vxitoQxy}
\end{equation}
Using eqn (\ref{vxitoQxy}) in the linearized equations for $\mathsf{Q}_1$ yields
\begin{eqnarray}
\beta_\pm&=&-\frac{\trp{1}}{\tau}\left(1+\frac{\pi ^2 K n^2}{A W^2}\right)+\frac{\lambda a}{2\eta(1+\dot{\gamma}^2  \tau ^2)}
\nonumber\\
&\pm&
\sqrt{\left[\frac{\lambda a}{2\eta(1+\dot{\gamma}^2\tau^2)}\right]^2-\dot{\gamma}^2\left(1+\frac{\lambda a\tau/\eta}{1+\dot{\gamma}^2\tau^2}\right)}.\label{betapm}
\end{eqnarray}
There are two modes. In the limit of a passive fluid, $a=0$, the modes collapse to a single mode corresponding to oscillations of the order parameter as it decays to its equilibrium value \abeds{given by eqn (\ref{eq:S})}: $\beta_\pm=-\lambda/\tau[1+\pi^2K/(AW^2)]\pm\mathrm{i}\dot{\gamma}$. Note the similarity between these damped oscillations and the damped oscillations in the startup problem, eqns~(\ref{Qxxstartup}) and (\ref{Qxystartup}).
A nonzero activity makes the two modes distinct.
In the limit of zero shear rate,  $\beta_-$ is negative and independent of activity even if $a\neq0$, and corresponds to the decay of the scalar order parameter of a passive isotropic nematic when it is perturbed from the isotropic value $S=0$. The other mode corresponds to the spontaneous flow and ordering of an active isotropic nematic when  $a>a_\mathrm{c}=[1+\pi^2K/(AW^2)]\eta/(\lambda\tau)$. Note that the confining channel walls raise the critical activity above the previously quoted critical value for unbounded space. The elastic constant $K$ only enters the growth rate if the channel width is finite.
\begin{figure}[t]
\centering
\includegraphics[height=2.4in]{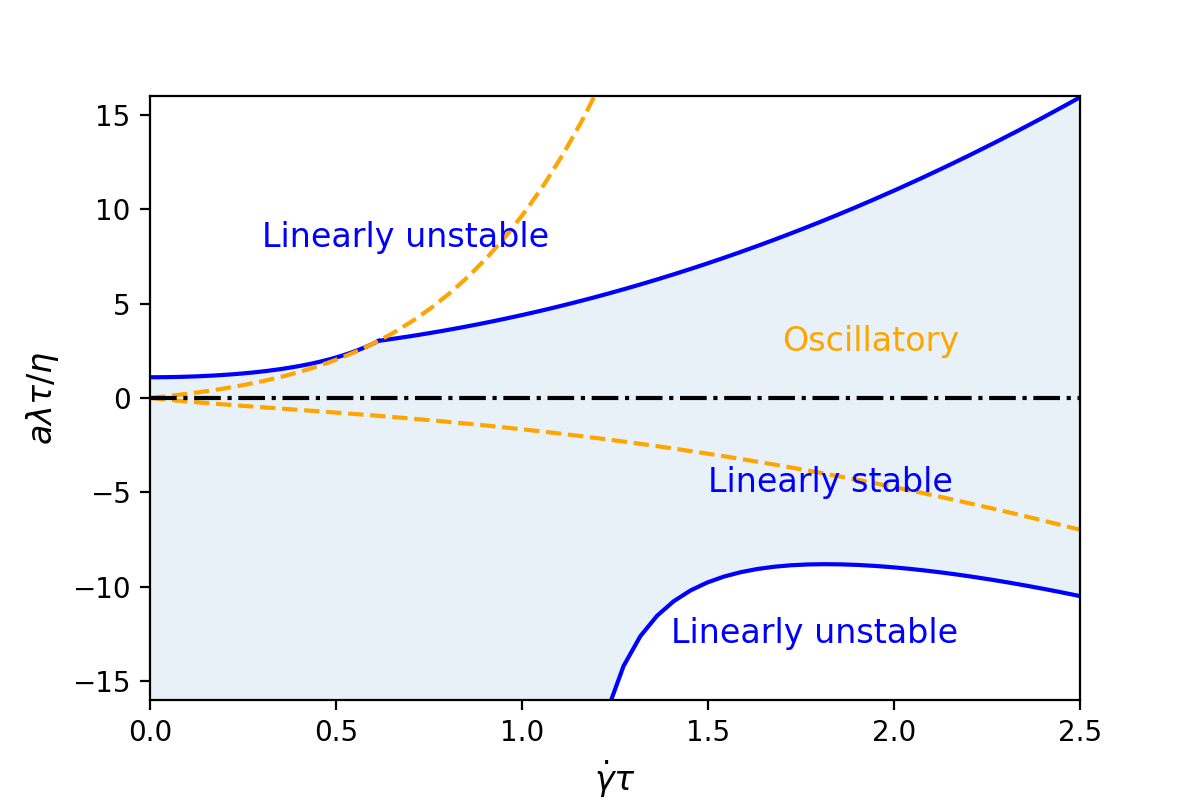}
\caption{
Linear stability analysis \trp{results} for a two-dimensional active gel in a straight channel of width $W$ subject to a shear flow with rate $\dot{\gamma}$. The Frank elasticity is small: $K=0.01 A W^2$. 
\trp{Simple shear flow is stable against perturbations in the shaded blue region,}
\trp{and the perturbations are oscillatory} in the region between two dashed lines. }
\label{fig:linearstability}
\end{figure}

In general, the critical activity for instability depends on the shear rate, and is found by determining when $\Re(\beta_+)=0$ for $n=1$. The modes are oscillatory when the square root in eqn~(\ref{betapm}) is imaginary, or when $a_-<a<a_+$, where
\begin{equation}
a_\pm=\frac{2\eta\dot{\gamma}}{\lambda}(1+\dot{\gamma}^2\tau^2)\left(\dot{\gamma}\tau\pm\sqrt{1+\dot{\gamma}^2\tau^2}\right).
\end{equation}
When $a_-<a<a_+$, the critical curve $\mathrm{Re}[\beta_+(n=1)]=0$ in the $\dot{\gamma}$-$a$ plane is given by 
\begin{equation}
a_\mathrm{1c}=2\frac{\eta}{\trp{\lambda}\tau}\left(1+ \dot{\gamma}^2 \tau^2 \right)\left(1+\pi^2\ell^2\right),
 \end{equation}
where $\ell$ is the dimensionless correlation length defined in the previous section. 
When $a<a_-$ or $a>a_+$, the growth rate is purely real, and the critical curve $\beta_+(n=1)=0$ is given by 
\begin{equation}
a_{2\rm{c}}=\frac{\eta}{\tau}\frac{\left(1+\dot{\gamma}^2 \tau^2\right)\left[(1+\pi^2\ell^2)^2+\dot{\gamma}^2\tau^2\right]}{\lambda(1+\pi^2\ell^2-\dot{\gamma}^2 \tau^2)}\label{a2c}
\end{equation}
Note that $a_{2\mathrm{c}}>0$ for $\sqrt{1+\pi^2\ell^2}>\dot{\gamma}\tau$, and $a_{2\mathrm{c}}<0$ for $\sqrt{1+\pi^2\ell^2}<\dot{\gamma}\tau$.


The stability boundaries are plotted in  Fig.~\ref{fig:linearstability} for the case of $\ell=0$ (zero Frank elasticity). The region of oscillatory growth rates, $a_-<a<a_+$, is the region between the dashed lines. The stable region is the shaded blue region between the solid blue curves, whereas the unstable regions are the white regions. Note that the upper stability boundary is given by $a_{1\mathrm{c}}$ in the oscillatory region, and $a_{2\mathrm{c}}$ in the non-oscillatory region. The lower stability boundary lies wholly in the non-oscillatory region, and is therefore given by $a_{2\mathrm{c}}$. Since the upper stability boundary near $\dot{\gamma}=0$  increases with shear rate, our results are in agreement with Muhuri et al.,\cite{MuhuriRaoRamaswamy2007} who found that shear counteracts the instability for extensile particles.
Surprisingly, we also find that shear can be \textit{destablilizing}  for contractile active particles if the magnitude of the activity is large enough.

\section{Straight channel: nonlinear spontaneous flows}\label{sec:nonlinear}


\trp{The linear analysis of the previous section predicts that simple shear flow with uniform nematic order is stable as long as the activity and externally imposed shear rate lie in the shaded region of Fig.~\ref{fig:linearstability}. However, there may be transitions to flow states that are not captured by linear stability analysis, and furthermore, the linear equations cannot describe the fully-developed flow states.} 
\trp{Thus,} we explore the activity-induced flow states and the transitions between them by numerically solving the full nonlinear equations, eqns (\ref{incompress})--(\ref{Qeqn}). 
We use the open source finite element software FEniCS\cite{Loggwells2010,LoggEtal_10_2012,AlnaesEtal2014} to solve the nonlinear equations, 
\trp{employing a} backwards Euler scheme 
to solve for the time dependence. 
We characterize 
\trp{t}he flow states by the 
\trp{spontaneous volumetric flow rate as well as the wall shear stress.}

The system is initialized with a small value of the nematic order parameter $S$, appropriate for an isotropic state. For sufficiently small values of the external shear, the direction of the activity-induced flow for $a>a_c$ 
depends on the 
\trp{configuration of the} nematic order. 
\trp{We can achieve positive flow---flow in the same direction the bottom wall moves---or negative flow---flow against the direction the bottom wall moves---by imposing appropriate initial conditions on the directors. These conditions will be described below for the extensile and contractile cases.}
\trp{The initial director fields also have small} random fluctuations.
Because we are neglecting inertial effects, we do not need to initialize the velocity field, which \trp{is}
determined from eqns (\ref{incompress})-(\ref{Qeqn}). 
\trp{Instead of attempting to simulate a very long channel,} we use periodic boundary conditions on the left and right boundaries of the channel. 
The length \trp{$L$} of the channel is chosen to be five times the width $W$\trp{;} 
we found \trp{this length} to be the longest channel \trp{length}
we could simulate in a reasonable amount of \trp{computing} time.  
\trp{We focus on situations in which the width $W$ of the channel is large compared to the correlation length $\sqrt{K/A}$ of the liquid crystal.} \trp{Therefore, our} simulations \trp{are carried out} with a small value of the Frank elasticity, $K/A=0.01 W^2$ (i.e. $\ell=0.1$)\trp{.} 
\trp{In our numerical calculations,} 
$W$ 
\trp{is} the unit of length, $\tau$ 
\trp{is} the unit of time, and $\eta/\tau$ \trp{is} 
the unit of pressure. We also define the dimensionless activity $\alpha=a{\lambda}\tau/\eta$,
\trp{and} restrict our simulations to the 
\trp{case} of needle-like particles, 
$\lambda=1$.
\begin{figure}[t]
\centering
\includegraphics[height=2.4in]{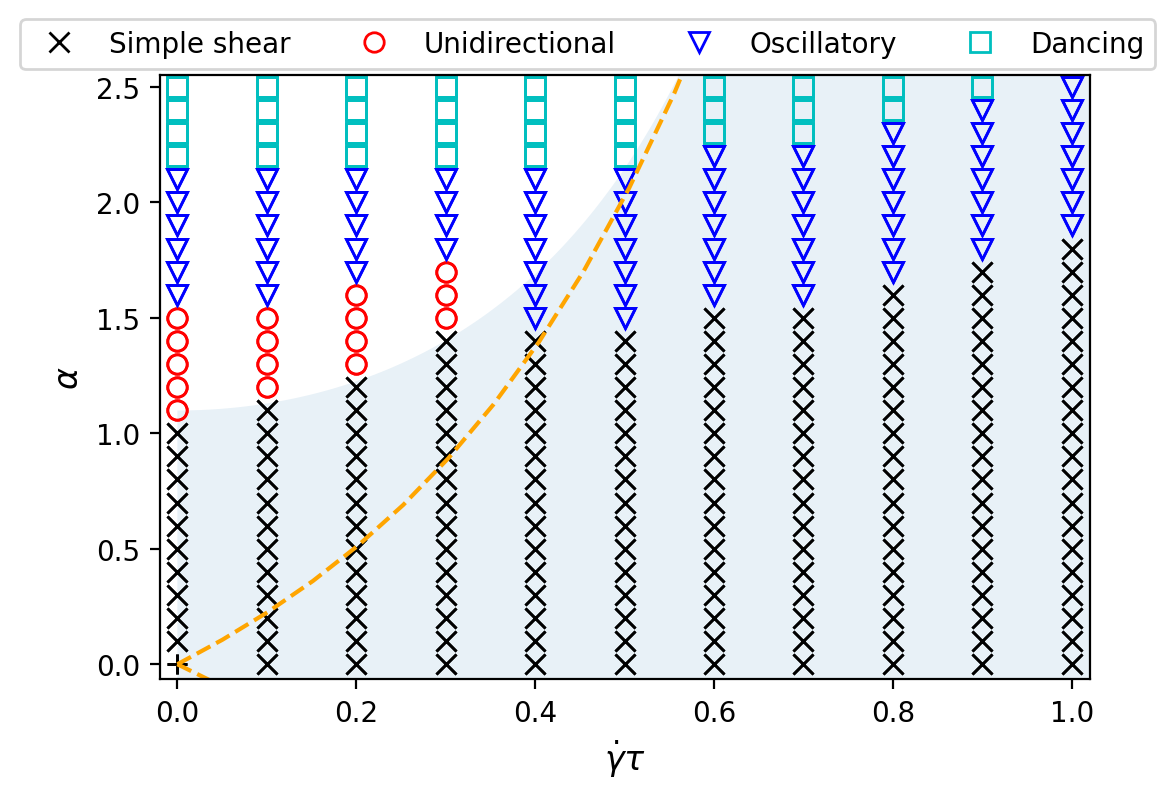}
\caption{
Numerically determined flow states for extensile fluids with $\ell=0.1$ and $\lambda=1$ in a straight channel. See the text for an explanation of how the flow states were determined. \trp{The blue shaded region and the dashed line indicate the linearly stable states and the boundary for oscillatory modes, respectively, for $\ell=0.1$ (compare with Fig.~\ref{fig:linearstability}).}
The \trp{location of the} transitions 
\trp{is} generally insensitive to whether \trp{the initial splay of the} nematic directors 
\trp{converges to the right (as in Fig.~\ref{fig:ext_unid}b) or the left (as in Fig.~\ref{fig:ext_unid}d).}
\trp{Note that depending on the noise in the initial conditions, some of states at higher $\alpha$ can either be dancing or oscillatory-like states; an example is shown in the SI.}
}
\label{fig:extensile}
\end{figure}

\begin{figure}[t]
\centering
\includegraphics[height=3.1in]{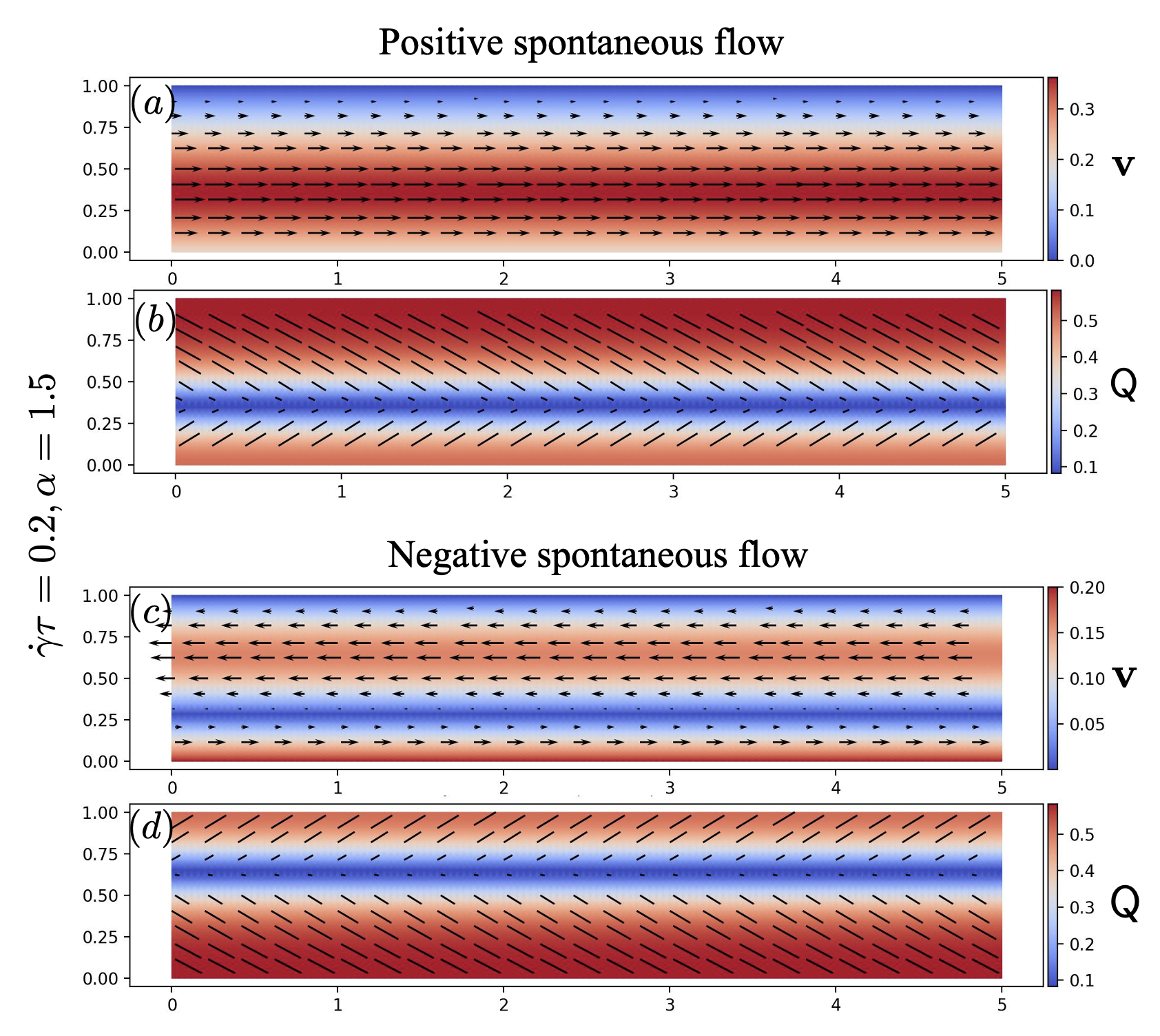}
\caption{
\trp{Steady states of spontaneous flow and nematic order for dimensionless activity $\alpha=1.5$ and dimensionless shear rate $\dot{\gamma}\tau=0.2$ (the bottom wall moves to the right, and the top wall is stationary).}
In the velocity field plots (panels (a) and (c)), color denotes flow speed, and arrows denote the direction of the flow. In the order parameter field (panels (b) and (d)), color denotes the scalar order parameter and lines denote the director field. \trp{Note that the leftward spontaneous flow in (c) is not sufficient to reverse the net flow near the moving wall, but leads to negative net flow near the stationary wall.}}
\label{fig:ext_unid}
\end{figure}

\subsection{Extensile fluids}

For extensile fluids, we find three types of flow states when the activity is above the critical value $a_c$: unidirectional, oscillatory, and dancing\trp{. These states are}  similar to three of the states found by Samui et al., \cite{samui2021flow} who studied an active nematic fluid confined to a channel in the absence of external shear. \trp{T}hese authors also found an active turbulent state at high activity, \trp{which we do not explore here}. \trp{The unidirectional flow is steady, consisting of a superposition of spontaneous flow and simple shear flow. The oscillatory flow is unsteady, with a pattern of flow and order that translates at a constant velocity along the channel, which makes the spatially-averaged wall stress constant in time. The dancing flow is truly unsteady, with a spatially-averaged wall stress that oscillates in time. These states will be described in more detail below.}
Fig.~\ref{fig:extensile} 
\trp{shows the phase diagram for} 
flow states for \trp{dimensionless} activity in the range $0 \leq \alpha \lesssim 2.5$ and shear rate in the range $0 \leq \dot{\gamma}\tau \lesssim 1$. 
\trp{To get positive spontaneous flow, we imposed initial conditions with the directors converging to the right, as in Fig.~\ref{fig:ext_unid}b. To get negative spontaneous flow, we imposed initial conditions with the directors converging to the left, as in Fig.~\ref{fig:ext_unid}d.}
We ran 
\trp{each} simulation until \trp{either all transients died out, or} $t=600\tau$, \trp{whichever came first}. \trp{The final state could either be a steady state or a state with regular periodic behaviour. Then we classified the states as follows}. 
\trp{The simple shear and unidirectional flow states} generally emerge at times $t\trp{<} 600\tau$. \trp{Both states are steady with negligible $y$-component of velocity, and these two flow states are easily distinguished since simple shear has the standard linear flow profile $v_x=\dot{\gamma}(W-y)$, whereas unidirectional flow has a spontaneous flow component added to the linear flow.} 
\trp{If there is a nonzero $y$-component of the velocity at the end of the simulation, we}
check 
\trp{for} oscillation\trp{s} 
in the average wall stress, 
\trp{$\bar{\sigma}_\mathrm{w}=\int_0^{L}\mathrm{d}x\sigma_{xy}(x,y=0)/L$},
for times in the range $550\tau$--$600\tau$. Negligible 
\trp{oscillation in the average wall stress} impl\trp{ies} an oscillatory flow state, while non-negligible values imply a dancing state. 
 Most of the points shown in Fig.~\ref{fig:extensile}
reached a steady or 
\trp{regular periodic} state by $t=600\tau$, or came very close to doing so.
\trp{But a few cases near transitions between flow states needed much longer to fully develop.} 



\begin{figure}[t]
\centering
\includegraphics[width=3.5in]{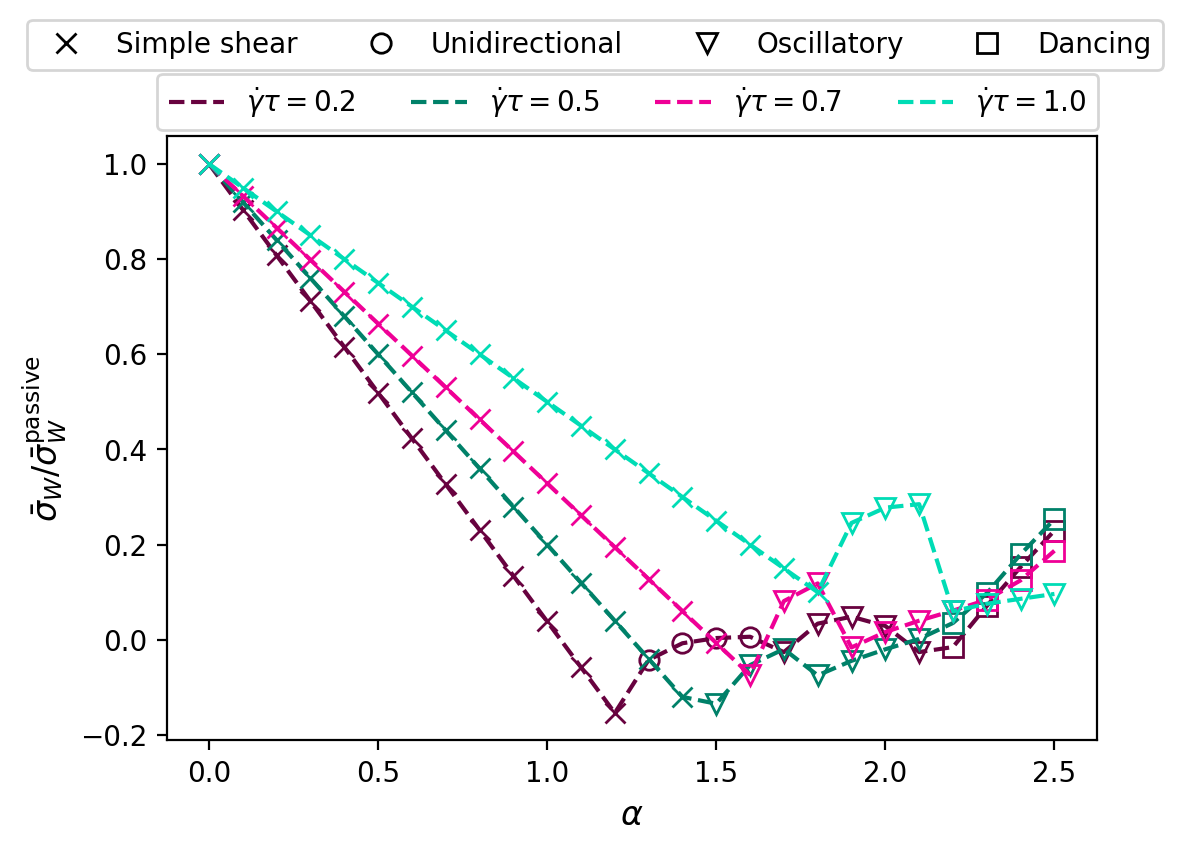}
\caption{
The \trp{spatially-}averaged wall stress  
on the bottom wall of a straight channel, normalized by the passive wall stress.
\trp{Because the average stress depends somewhat on the spatial period of the oscillatory and dancing flow states, only the results for one representative period are shown.}
For 
dancing flow \trp{states}, \trp{the spatially-averaged wall stress oscillates in time, and} the square symbols show the \trp{time-}average\trp{d} value of the 
wall shear stress.
}
\label{fig:extensilewall}
\end{figure}

The \trp{limit of stability for the simple shear flow states} 
in the Fig. \ref{fig:extensile} is the boundary between the 
\trp{region with black crosses and the regions with other symbols.} 
We observe \trp{that} the numerical \trp{limit of} stability 
\trp{for simple shear flow} matches very well
with the \trp{prediction of} linear stability analysis (filled blue region), but only for the transition from the simple shear to unidirectional flow\trp{, $\dot{\gamma}\tau\lesssim0.3$}. The disagreement between the linear stability boundary and the transition from simple shear flow to oscillatory flow 
\trp{may be} due to \trp{our neglect of the possibility that the perturbation could depend on $x$ as well as $y$.} 
In the \trp{region of simple shear flow (black crosses in Fig.~\ref{fig:extensile}), our numerical results show that the wall stress decreases with activity, in agreement with eqn~ \ref{wallshear}. Fig.~\ref{fig:extensilewall} shows the numerically computed wall stress, normalized by the passive (viscous) stress. When the flow state is simple shear, activity reduces the total wall stress in proportion to the activity, in accord with the general understanding that extensile particles with activity reduces the effective viscosity.~\cite{hatwalne04}}

\begin{figure}[t]
\centering
\includegraphics[height=2.3in]{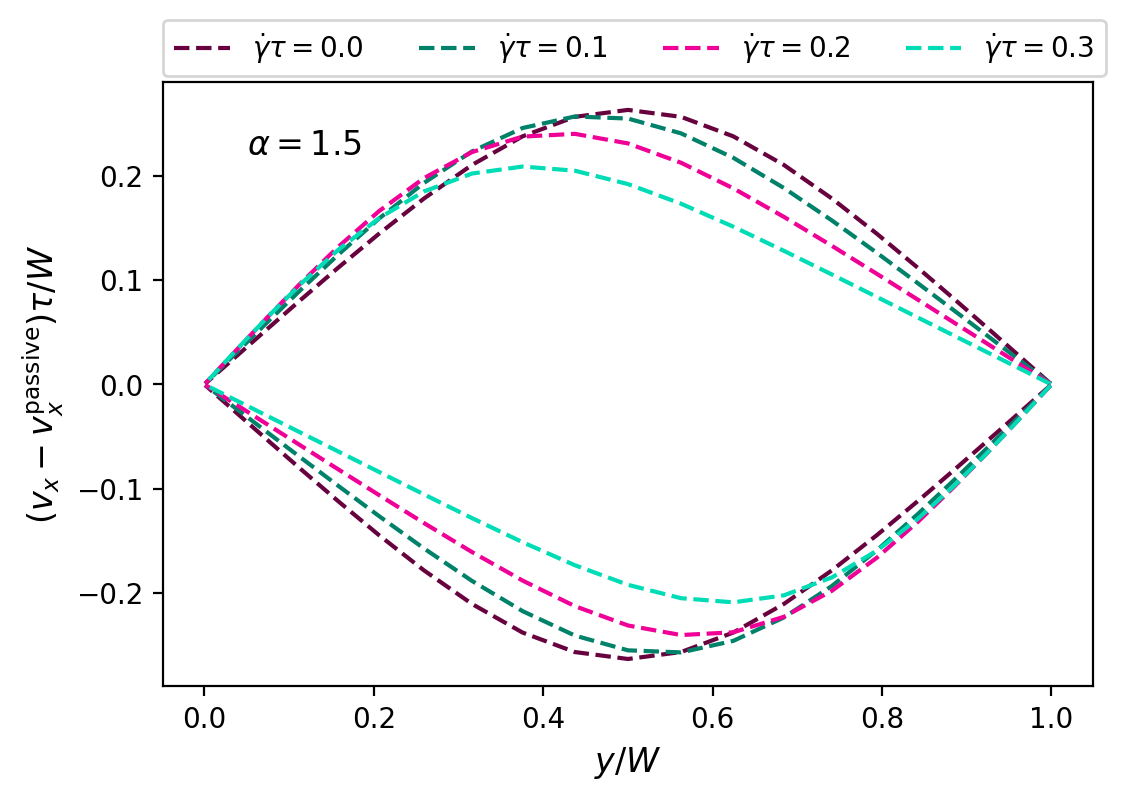}
\caption{
Spontaneous component of velocity profiles 
in dimensionless units for 
unidirectional flow 
at different rates of 
externally imposed shear. For each value of the shear rate, there are two branches, with the upper branch corresponding to the positive spontaneous flow, and the lower branch corresponding to the negative spontaneous flow. 
}
\label{fig:extvx_u0}
\end{figure}

\textbf{Unidirectional flow.} When the externally imposed shear is \trp{in the range $0\le\dot{\gamma}\tau\lesssim0.3$, and the dimensionless activity is in a relatively narrow band near $\alpha\approx1$ (Fig.~\ref{fig:extensile}),}
activity creates a steady 
unidirectional flow along the $x$-axis~\trp{(Fig.~\ref{fig:ext_unid})}. \trp{The activity-induced component spontaneously breaks the left-right symmetry of the channel, with the actual direction of the active flow component determined not by the imposed external shear but instead by the initial conditions of the directors, as described above.}
\trp{Since the total shear rate vanishes at the value of $y$ at which the flow rate has an extremum, the scalar parameter vanishes at this same value of $y$ (Fig.~\ref{fig:ext_unid}).}
\trp{Fig.~\ref{fig:extvx_u0}} shows the flow profile subtracting off the imposed shear flow for fixed activity and various values of $\dot{\gamma}$ for both the left-moving and right-moving spontaneous flows.
It indicates that the \trp{spontaneous active component of the flow depends on $\dot{\gamma}$; in other words, the}
total flow is not 
simply a 
superposition of the passive 
\trp{shear flow $v_x=\dot{\gamma}(W-y)$} and \trp{the spontaneous flow at} \emph{zero} 
externally imposed shear. 

\begin{figure}[t]
\centering
\includegraphics[height=2.4in]{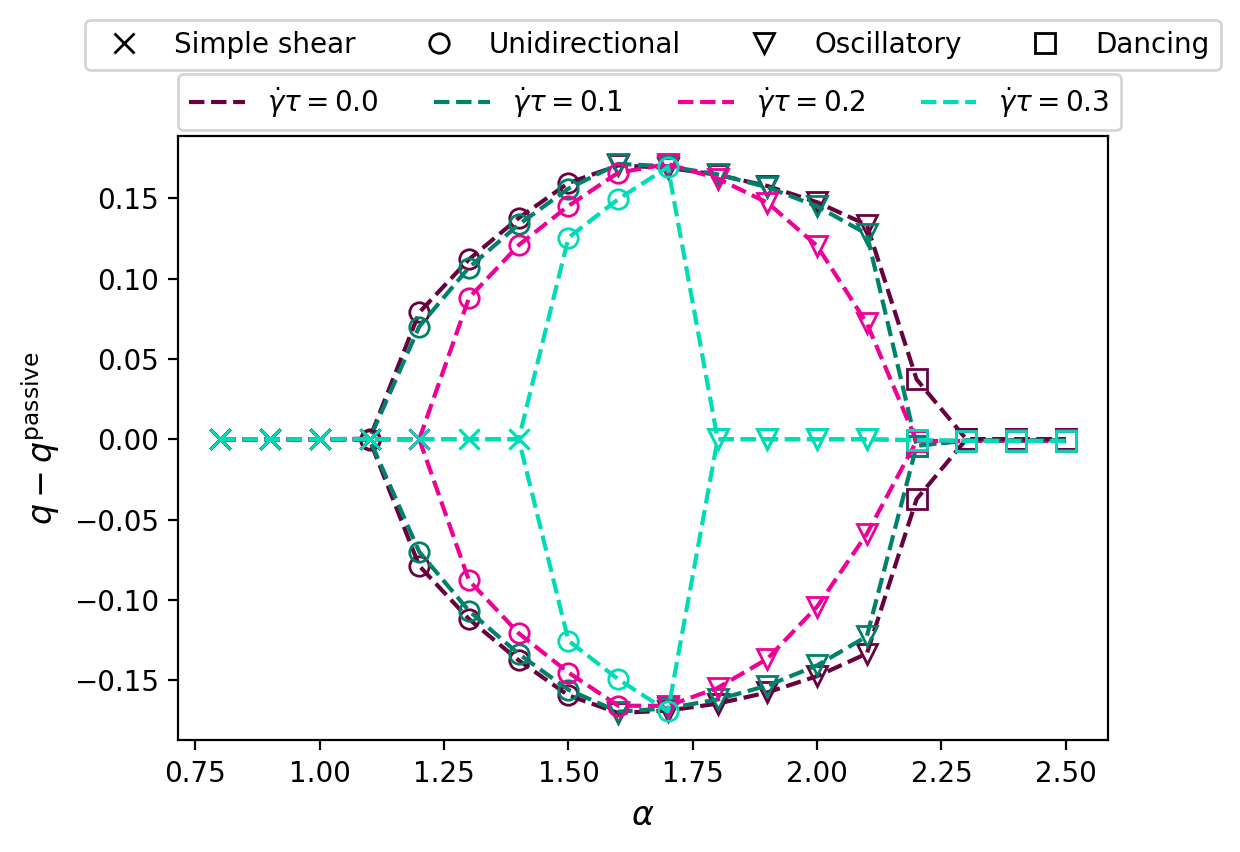}
\caption{
Dimensionless activity-driven 
\trp{volumetric flow rate} 
\trp{versus dimensionless} activity in a straight channel. 
The symbols denote the flow states and the colors denote the externally imposed shear rate. 
\wl{The wavelength is $5W/4$ for all cases with nonzero activity-driven flow rate shown in the figure.}
}
\label{fig:extensileflux}
\end{figure}

\begin{figure}[t]
\centering
\includegraphics[height=4.3in]{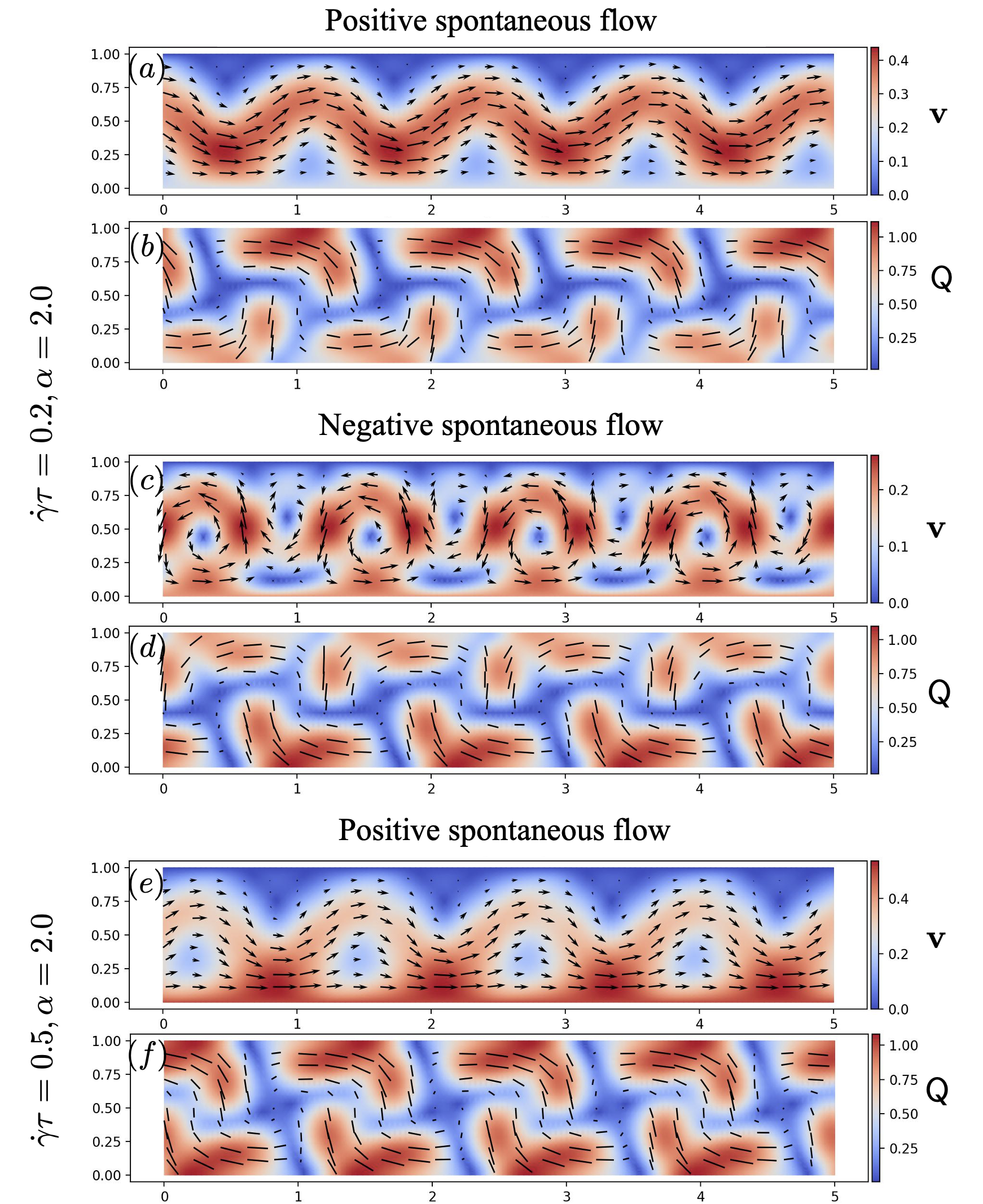}
\caption{
Examples of the unsteady oscillatory flow state at $t=600\tau$. 
\trp{Panels} 
(a), (b), (c) and (d) show the flow patterns \trp{and order parameter fields} corresponding to positive and negative spontaneous flows 
\trp{for} small external shear rate. Panels (e) and (f) show 
a case with 
larger 
\trp{shear rate.} At this shear rate, we only observe positive spontaneous \trp{flow}. 
}
\label{fig:ext_trans}
\end{figure}

\trp{To better characterize these flows, we subtract the passive volumetric flow rate  from the total volumetric flow rate to get the dimensionless activity-induced volumetric flow rate (per unit channel width), 
\begin{equation}
    q^\mathrm{active}\equiv\left(\int_0^W\mathrm{d}y v_x-\frac{\dot{\gamma}W^2}{2}\right)\frac{\tau}{W^2},
\end{equation}
shown in Fig.~\ref{fig:extensileflux}. This quantity serves as an order parameter describing the transitions among the various flow states. Fig.~\ref{fig:extensileflux} shows} that the activity-driven 
\trp{flow rate has the same} magnitude \trp{for the left-moving and right-moving flows, and also that the amplitude of the unidirectional flows increases as the activity increases.} 

\trp{Examining Fig.~\ref{fig:extensilewall} for the case of $\dot{\gamma}\tau=0.2$ reveals that reduction of the \wl{normalized} wall stress with increasing activity ceases at the onset of the unidirectional flow, and the \wl{normalized} wall stress at $y=0$ starts to increase slightly as activity increases further. The active component of the wall stress at $y=0$ in the unidirectional flow has the opposite sign compared to that of the simple shear flow, as can be seen from the opposite orientation of the directors near the wall $y=0$ in Fig.~\ref{fig:ext_unid}b and Fig.~\ref{fig:eg}. Also, the active component of the flow changes the sign of the flow gradient near the wall, as can be seen from Fig.~\ref{fig:extvx_u0}. These two effects together lead to the rise in \wl{the normalized} wall stress at the onset of unidirectional flow.}

\textbf{Oscillatory flow.} \trp{Our phase diagram of flow states shows that for $\dot{\gamma}\tau\lesssim0.3$, there is a transition with increasing activity from the unidirectional flow states to two-dimensional oscillatory flows (Fig.~\ref{fig:ext_trans}). 
When $\dot{\gamma}\tau\gtrsim0.3$, the simple shear states transition directly to two-dimensional oscillatory flows as activity increases.}
\trp{Although the oscillatory flow states are unsteady, with the velocity and order parameter taking the form of a  traveling wave,}  
the flow pattern and order parameter configuration rigidly translate in the $x$ direction 
\trp{with wave speed} $v_{\mathrm T}$. \trp{In other words, in the frame moving relative to the channel walls with speed $v_\mathrm{T},$ the streamlines meander in space but are steady. Likewise, the configuration of the order parameter tensor is steady in this frame. 
Because we use periodic boundary conditions, the flow field and orientational order parameter must have period in $x$ equal to the channel length $L$. But these fields could also have a shorter period, which must evenly divide the total channel length. Since we use a channel length $L=5W$, the possible wavelengths for a  periodic configuration are $5W$, $5W/2$, $5W/3,$ .... Different wavelengths are selected in the dynamical final state depending on 
\wl{the initial state of the nematic directors}, as well as the value of the activity and the imposed shear.}
Because it is difficult to determine the relationship between the 
\wl{random fluctuations imposed on the initial directors} and the wavelength that is finally selected, we did not make a systematic study of all the possible wavelengths.
It is natural to worry that the steady translation of the flow field and order parameter pattern could be an artifact of the periodic boundary conditions. In Sec.~\ref{nonlinear annular}, we study an annular geometry as a single domain without the need for periodic boundary conditions. 
Since we also observe an oscillatory flow state with constant angular wave speed in that situation, 
we are confident the constant wave speed $v_T$ we see in the straight channel is not an artifact of the period boundary conditions.

We measured the 
\trp{volumetric flux for} 
times in the range $t=550$--$600\tau$, which is 
\trp{when} the system is generally in its final dynamically stable state. In the final state, the 
\trp{volumetric flow rate} and wall shear stress of the oscillatory flows are 
constant.
 For small externally imposed shear (e.g. \wl{$\dot{\gamma}\tau \leq 0.2$} in Fig.~\ref{fig:extensile}), the spontaneous activity-induced flows can be either positive or negative\trp{, depending on the form of the splay in the initial conditions for nematic order, as for the unidirectional flows.} See Figs.~\ref{fig:ext_trans}a--d (movies are in the SI).
 For positive spontaneous flow, the streamlines undulate, but the externally imposed shear breaks the up-down symmetry of the waves with respect to the horizontal centerline of the channel. The velocity at the valleys of the waves is higher than at the peaks. For negative spontaneous flow, since the activity-induced flow is opposite 
 to the \trp{direction of the} externally imposed shear \trp{flow}, the flow more easily forms circular streamlines. Thus, for $\dot{\gamma}\neq0$, the absolute value of the activity-driven flux of negative spontaneous flows 
 is slight\trp{ly} smaller than the flux for the positive spontaneous flows, \trp{as can be seen by looking very closely at Fig.~\ref{fig:extensileflux}}. 

\begin{figure}[t]
\centering
\includegraphics[height=2.3in]{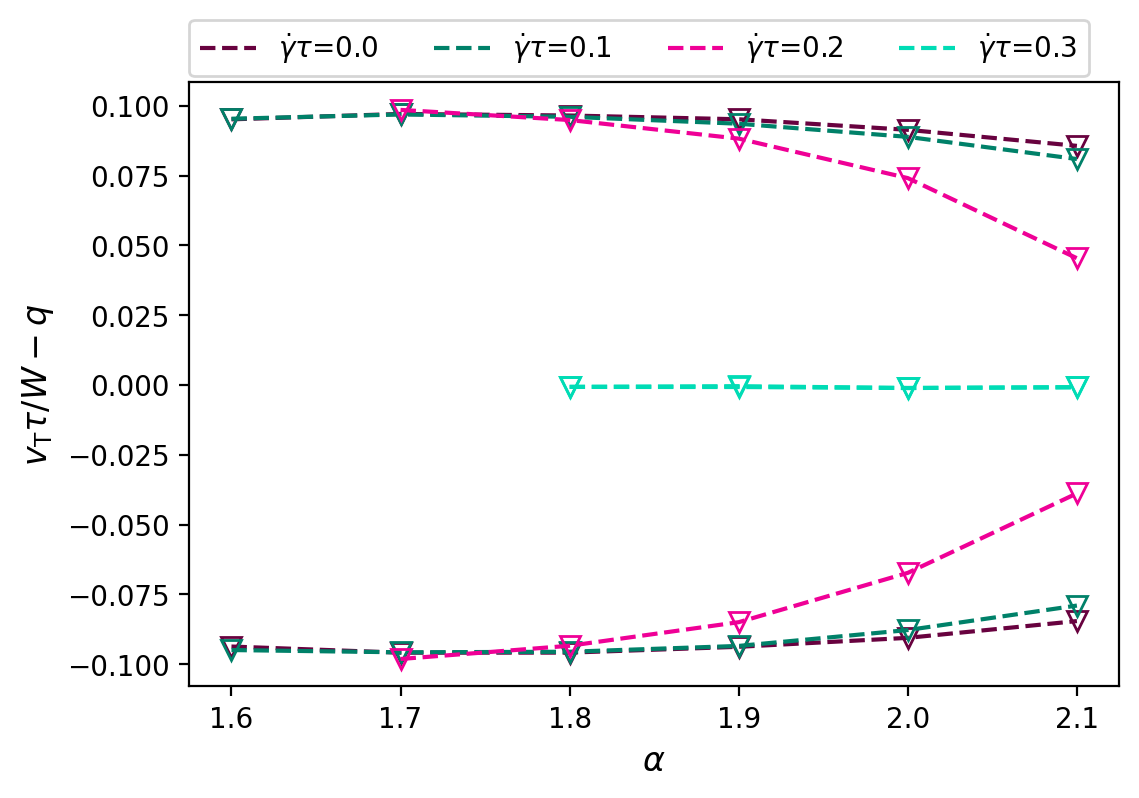}
\caption{
\trp{Wave} speed $v_\trp{T}$ of the flow pattern 
\trp{relative to the average dimensionless flow rate for} the oscillatory flow state with various (small) externally imposed shear rates. 
}
\label{fig:trans_speed}
\end{figure}

 The direction of the spontaneous flow not only 
 \trp{determines} the shape of the streamlines, 
 but also determines the direction of translation of the total flow pattern, 
 including the passive viscous flow.  
 For positive spontaneous flow, the total flow pattern translates in the $+x$ direction, while for the negative spontaneous flow case, it translates in the $-x$ direction. The activity-driven 
 \trp{volumetric flow rate} is nonzero but generally decreases with increasing activity as shown in Fig.~\ref{fig:extensileflux}. 
 Fig.~\ref{fig:trans_speed} 
 shows that the 
 wave speed $v_{\rm{T}}$ 
pattern is faster than the flux, and the difference between these two quantities decreases with the growth of the activity.

We now turn to 
\wl{larger externally imposed shear (e.g. $\dot{\gamma}\tau \geq 0.3$ in Fig.~\ref{fig:extensile})}. 
\trp{In this case,} only the positive spontaneous flow appears; the symmetry is broken by the flow imposed by the external shear. The activity-driven \trp{volumetric flow rate is} 
zero 
because 
\trp{the imposed shear rate} is large enough to close the streamlines. Interestingly, \trp{our numerical results indicate} 
that the 
\trp{wave} speed is equal to the 
\trp{average volumetric flow rate of simple shear, $v_\mathrm{T}=\dot{\gamma}W/2$.}

\begin{figure}[t]
\centering
\includegraphics[height=1.4in]{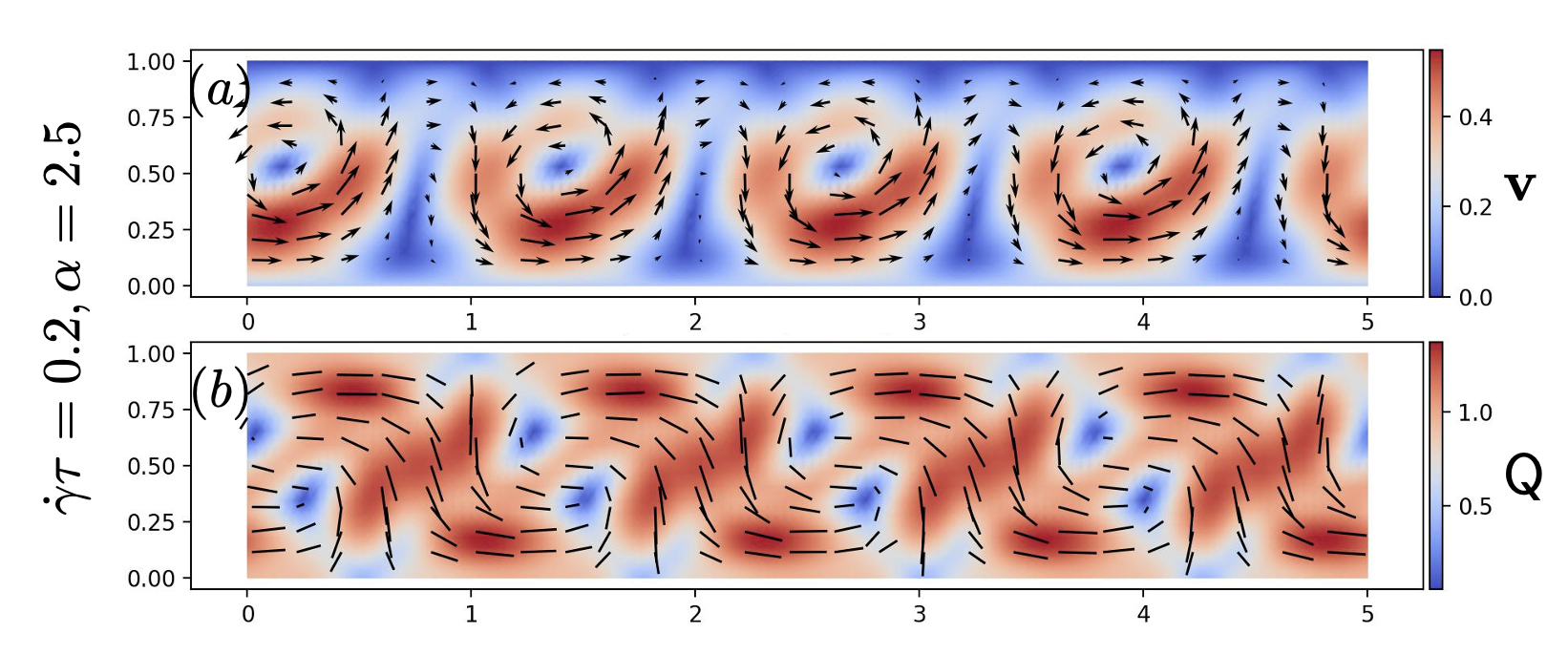}
\caption{
\trp{Snapshot} of \trp{a representative} 
unsteady dancing flow state at $t={599.8}\tau$. 
In the velocity field plots, colors denote 
flow speed,
and arrows denote 
flow direction. In the order parameter field, colors denote the scalar order parameter, 
and lines denote \trp{directors}. 
}
\label{fig:ext_dan}
\end{figure}

\textbf{Dancing flow.}
At higher activity\trp{, the flow field and tensor order parameter field become unsteady in any frame, and we find states {(Fig. \ref{fig:ext_dan}; movies are in the SI)} analogous to the dancing flows found by {Shendruk et al.~\cite{shendruk2017dancing}} and Samui et al.~\cite{samui2021flow} in their study of active nematic flow in a two-dimensional channel}.
The volumetric flow rate of dancing flow is still constant with time. Additionally, in the range we study ($\alpha \leq 2.5)$, 
when activity 
is large enough to dynamically close all streamlines for the part of the flow that is activity-driven, the total flux is the same as in the passive case. 
\trp{As in the case of the oscillatory flows, sometimes we find multiple states at the same values of parameters. For example, n}oise in the initial conditions may cause the system to exhibit oscillatory-like states  \trp{in the region of the phase diagram where dancing flows are also found.}

\trp{Given a director configuration $\hat{\mathbf{n}}=\cos\phi\hat{\mathbf{x}}+\sin\phi\hat{\mathbf{y}}$, we may define the topological charge inside a closed loop by computing $\int\mathrm{d}\phi=2\pi m$ around the loop, where $m$ is the charge. Applying this definition to the configuration in Fig. \ref{fig:ext_dan}b may be problematic because the order parameter $S$ vanishes not just in small cores but in extended two-dimensional regions. If the loop drawn to encircle a potential topological defect crosses a region where $S$ vanishes, the angle $\phi$ and the topological charge are ill-defined. Nevertheless, we can simply look at the director configuration of dancing flow and see that there are parts of the configuration around the regions of small $S$ near the center of the channel that closely approximate the director field of $+1/2$ defects.}
The 
$+1/2$ defects appear 
in pairs\trp{,} and the two defect cores move with undulations \trp{of the flow} in opposite directions 
leading to \trp{the pairs} exchang\trp{ing} 
\trp{partners with the pair to the immediate left and immediate right,} consistent with the Ceilidh dance observed by Shendruk ~\cite{shendruk2017dancing} and Samui \cite{samui2021flow}.

\begin{figure}[t]
\centering
\includegraphics[height=2.4in]{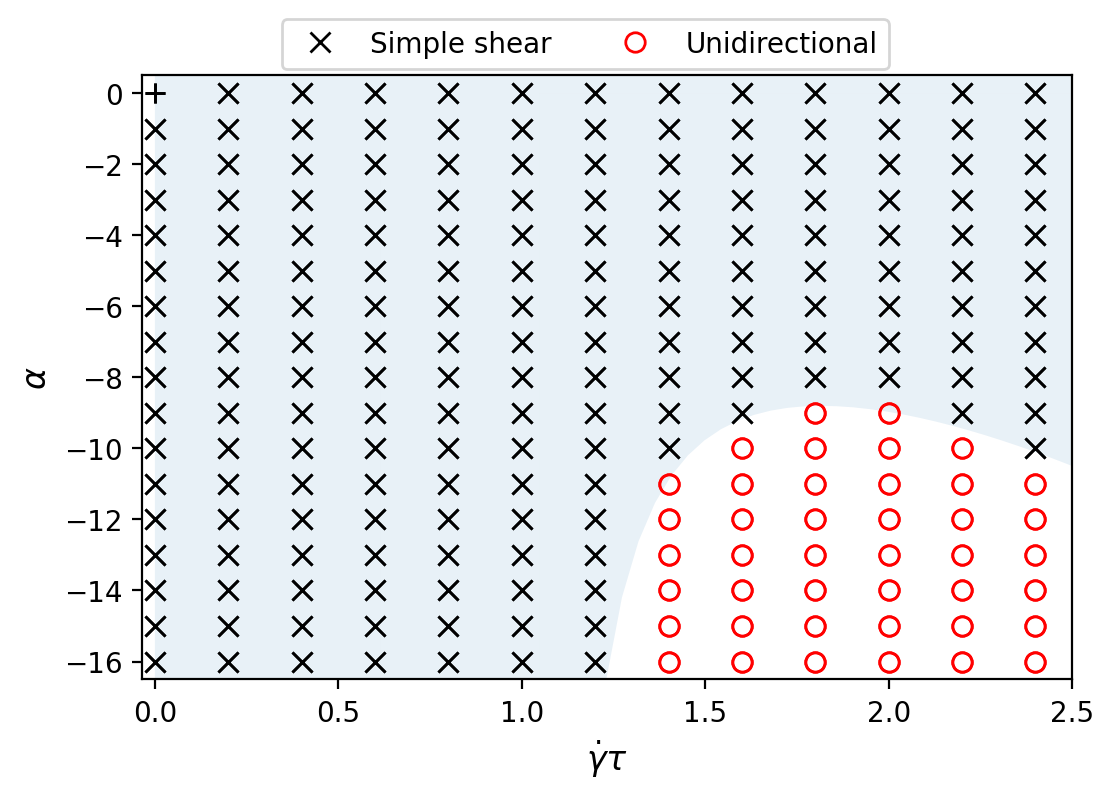}
\caption{
Flow states for contractile \trp{particles} 
in \trp{a} 
straight channel \trp{with} $\ell=0.1$. 
\trp{As in Fig.~\ref{fig:extensile},} the blue \trp{shaded} region is linearly stable, and the \trp{modes of the linearized equations are damped but oscillatory above the} dashed line.
The  
\trp{finite element results are} 
insensitive to the initial conditions of the director field.
}
\label{fig:contractile}
\end{figure}
\begin{figure}[t]
\centering
\includegraphics[height=3.1in]{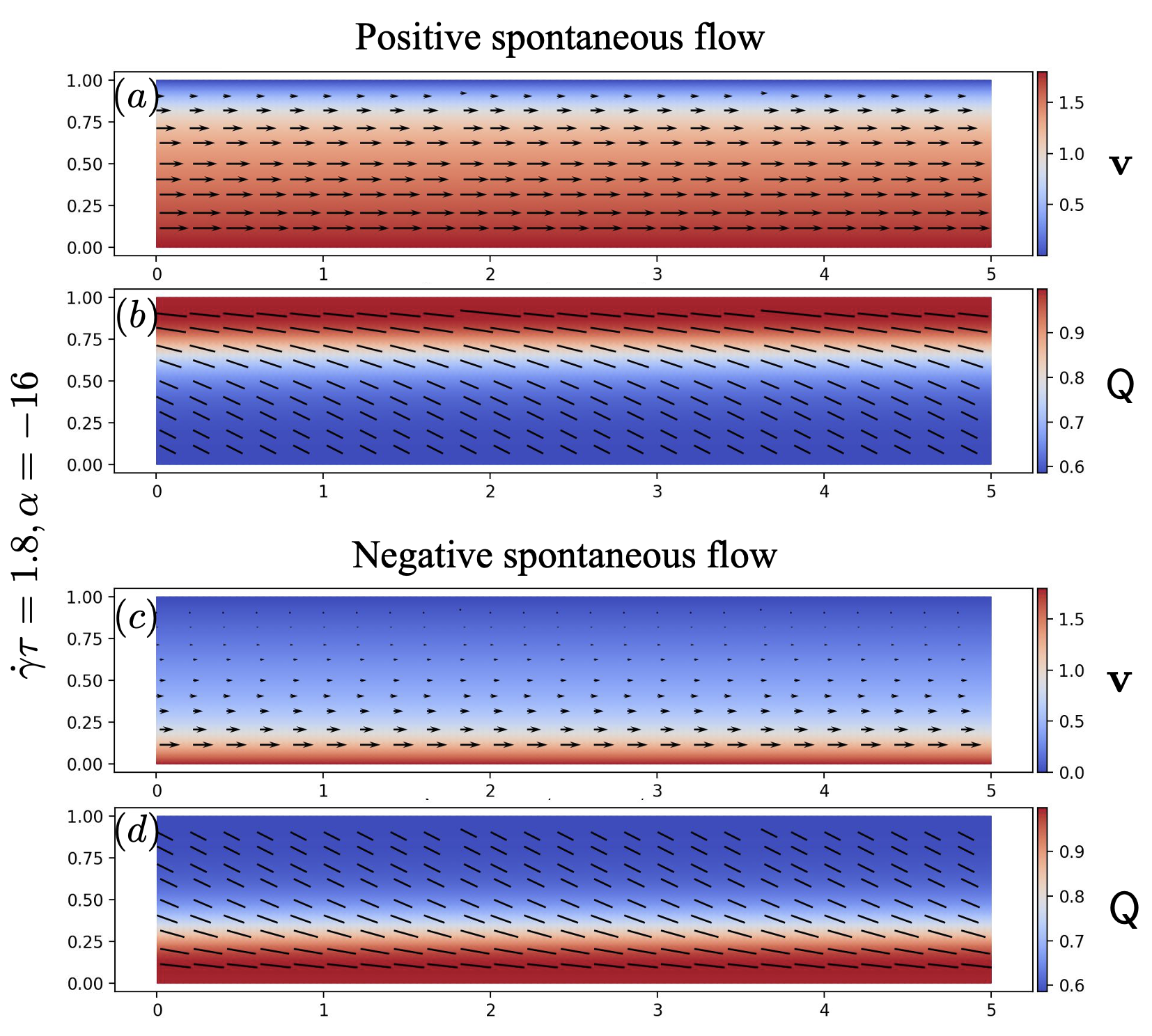}
\caption{
An example of the velocity and order parameter fields of positive and negative spontaneous \trp{flow} 
\trp{for} contractile \trp{particles.} 
In the velocity field plots 
\trp{(a) and (c)}, color denotes the 
flow speed,
and arrows denote flow direction. In the order parameter field plots \trp{(b) and (d),} 
color denotes the scalar order parameter, and lines denote \trp{the directors.} 
}
\label{fig:con_unid}
\end{figure}

The \trp{spatially averaged} shear stress imposed by the active flow on the moving wall also oscillates in time. 
The average wall shear stress no longer decreases linearly with activity in the spontaneous flow region. 


\subsection{Contractile fluids}
\trp{Negative activity corresponds to contractile particles. When the activity is sufficiently negative and the shear rate is large enough,} $\dot{\gamma}\tau>\sqrt{1+\pi^2\ell^2}$ \trp{, we observe unidirectional flow states in our finite-element calculations. The stability boundary that we find in our numerical calculations is}
\trp{c}onsistent with the \trp{results of our} linear stability analysis (Fig.~\ref{fig:contractile}).
\trp{As in the extensile case, we get {both} positive 
\trp{and} negative flows, \trp{depending on whether the initial configuration of the directors bends downward as in Fig.~\ref{fig:con_unid}b, or upward as in Fig. \ref{fig:con_unid}d.}}
\trp{After transients have died out, the active component of the volumetric flow rate is equal in magnitude for the positive and negative flows, and the amplitude of the flow rate increases as the magnitude of the activity increases.} \trp{It is well-known that contractile elongated particles in a shear flow enhance the shear viscosity.~\cite{hatwalne04} Thus, the wall stress (normalized by passive stress) increases linearly with the magnitude of the activity when the flow is simple shear, according to eqn~\ref{wallshear}. 
When the flow transitions to unidirectional flow, we also find that the \wl{normalized} wall \wl{stress} increases linearly with the magnitude of the activity, however with a slightly smaller \wl{absolute value of} slope.}
The figures showing the dependence of the active component of flow rate and the dependence of the \wl{normalized} wall \wl{stress} on activity are in the SI.

\begin{figure}[t]
\centering
\includegraphics[height=2.4in]{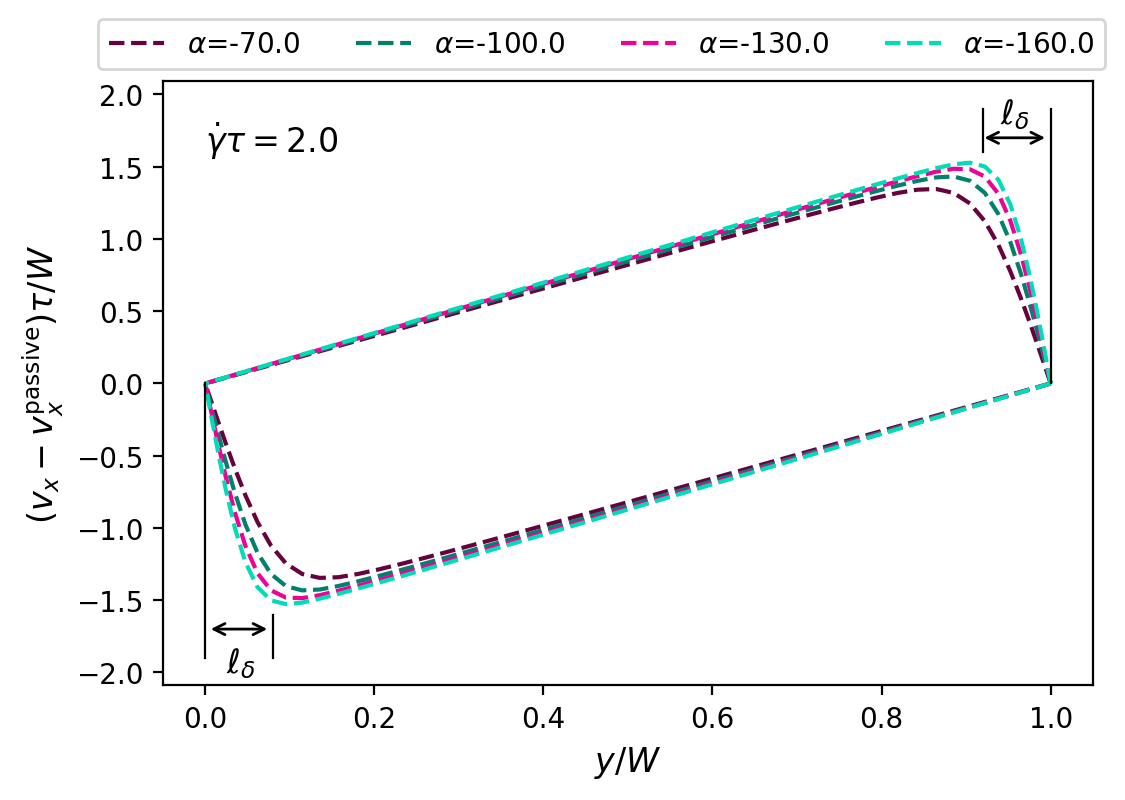}
\caption{
Active component of the velocity profile 
in dimensionless units for \trp{contractile particles in the} unidirectional flow state, for $\dot{\gamma}\tau=2.0$ and various activities. 
The upper branches correspond to positive spontaneous flow, and the lower branches correspond to negative spontaneous flow. The boundary layer thickness is denoted by $\ell_\delta$.
}
\label{fig:cont_u0}
\end{figure}

\begin{figure}[t]
\centering
\includegraphics[height=2.4in]{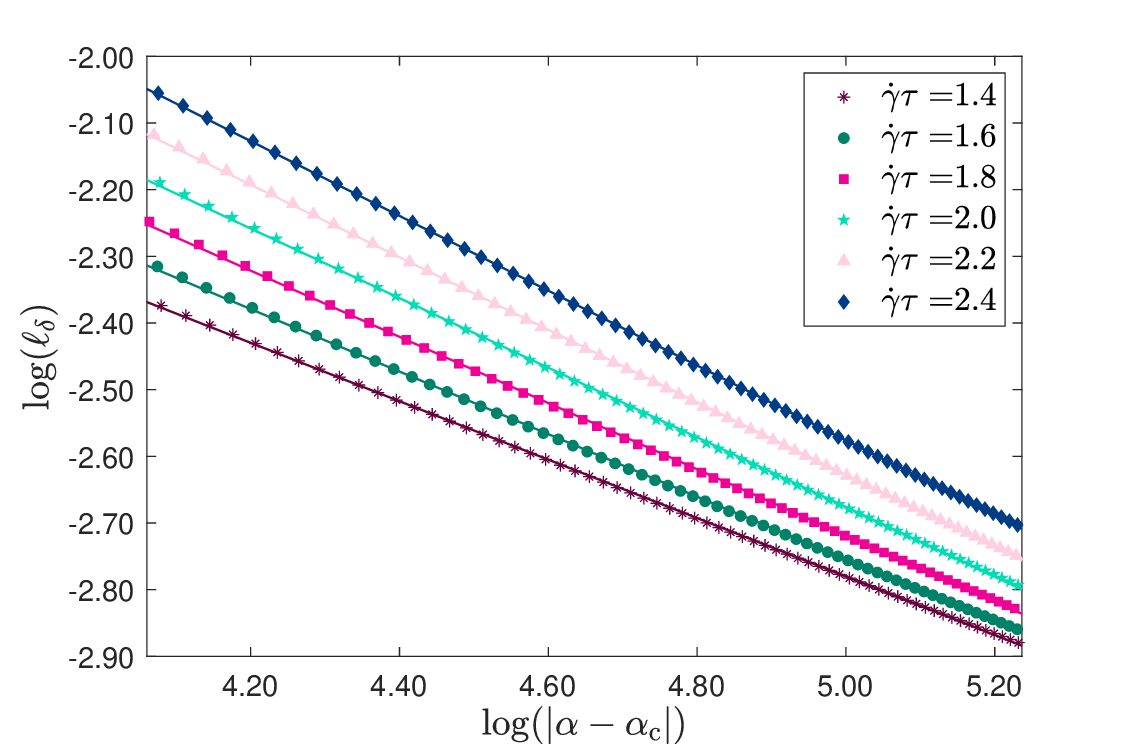}
\caption{
{Log-log plot of dimensionless boundary layer thickness $\ell_\delta$ vs. $\alpha-\alpha_c$, 
for various shear rates (legend).
These results indicate that $\ell_{\delta}\propto|\alpha-\alpha_c|^\zeta$, where $\zeta$ gradually changes from $-0.44$ to $-0.56$ with increasing $\dot{\gamma}\tau$.}}
\label{fig:Blayer}
\end{figure}

When the magnitude of the activity becomes large, we observe a boundary layer in the flow velocity. Since we found only  steady-state unidirectional flow states for contractile activity, it is computationally more efficient to reduce the governing partial differential equations to ordinary differential equations [see eqns~(\ref{vxstdy}--\ref{Qxystdy}) below] and solve them using the bvp5c solver of MATLAB.\cite{kierzenka2008bvp}
 Fig.~\ref{fig:cont_u0} shows the active component of the flow 
 for the positive and negative spontaneous flows of contractile \trp{gels.} 
When the absolute value of the activity is large, we observe that the spontaneous component of the flow approaches simple shear flow, with a boundary layer of \wl{dimensionless} thickness $\ell_\delta$ near one of the walls, which we define as the displacement boundary layer thickness~\cite{KunduCohen2008} 
 \begin{equation}
     \ell_{\delta}\equiv\frac{\int_0^W \mathrm{d} y \left(\dot{\gamma}_0 y-(v_x-v_x^{\rm{passive}})\right) }{\int_0^W \mathrm{d} y \dot{\gamma}_0 y},
 \end{equation}
 where $\dot{\gamma}_0=\mathrm{d} (v_x-v_x^{\rm{passive}})/{\mathrm{d}y}$ at $y=0$ for positive spontaneous flow. The boundary layer thickness is the same for positive and negative spontaneous flow.
 Fig.~\ref{fig:cont_u0} shows that the peak flow speed of the active component is higher and the boundary layer is thinner for larger magnitudes of the activity. {From Fig. \ref{fig:Blayer},} we find that $\ell_\delta\propto|\alpha-\alpha_c|^\zeta$, where $\zeta$ is close to \wl{$-0.5$}, but 
 \wl{its magnitude} increases with $\dot{\gamma}\tau$. This dependence will be studied in another publication.

\subsection{Weakly nonlinear analysis for $\dot{\gamma}=0$.}
\label{weaknonlinear}
To conclude this section, we turn to a weakly nonlinear analysis of the spontaneous steady unidirectional flow near the transition from the motionless isotropic state.\cite{OhmShelley2022} We {continue to assume $\lambda=1$ and} only consider the case of zero shear rate, $\dot{\gamma}=0$, leaving the case of nonzero $\dot{\gamma}$ for another publication. Assuming that the velocity field, order parameter tensor, and pressure depend only on the coordinate $y$, the dimensionless governing equations are 
\begin{eqnarray}
v_x''-\alpha Q_{xy}'&=&0\label{vxstdy}\\
-p'+\alpha Q_{xx}'&=&0\label{vystdy}\\
\ell^2Q_{xx}^{\prime\prime}-Q_{xx}+v_x' Q_{xy}&=&0\label{Qxxstdy}\\
\ell^2Q_{xy}^{\prime\prime}-Q_{xy}-v_x'Q_{xx}+
v_x'&=&0\label{Qxystdy}
\end{eqnarray}
with no-slip boundary conditions $v_x(0)=v_x(1)=0$ and  no-torque (Neumann) boundary conditions $Q_{ij}'(0)=Q_{ij}'(1)=0$. The prime denotes a derivative with respect to $y$. We already saw in Sec.~\ref{stability} that the motionless, distorted state at zero imposed shear rate is unstable when $\alpha>\alpha_\mathrm{c}$, where $\alpha_c=(1+\pi^2\ell^2)
$ is the dimensionless critical activity. Here we study the spontaneous flow and weak ordering for $\alpha=\alpha_\mathrm{c}+\delta\alpha$, with $\delta\alpha>0$. Assuming the balance $Q_{xx}\approx v_x^\prime Q_{xy}$ in eqn~(\ref{Qxxstdy}) suggests that to leading order, $v_x=\mathcal{O}(\delta\alpha^{1/2})$, $Q_{xy}=\mathcal{O}(\delta\alpha^{1/2})$, and $Q_{xx}=\mathcal{O}(\delta\alpha)$. Thus, we expand in powers of $\delta\alpha^{1/2}$:
\begin{eqnarray}
    v_x&=&\delta\alpha^{1/2}v_x^{(1)}+\delta\alpha v_x^{(2)}+\delta\alpha^{3/2} v_x^{(3)}+\dots\\
    Q_{ij}&=&\delta\alpha^{1/2}Q_{ij}^{(1)}+\delta\alpha Q_{ij}^{(2)}+\delta\alpha^{3/2} Q_{ij}^{(3)}+\dots. 
\end{eqnarray}
At $\mathcal{O}(\delta\alpha^{1/2})$, we find the steady versions of the linearized equations we used in Sec.~\ref{stability} to solve for the growth rate,
\begin{eqnarray}
    v_x^{(1)\prime\prime}-\alpha_\mathrm{c}Q^{(1)\prime}_{xy}&=&0\label{v1eqn}\\
    \ell^2Q^{(1)\prime\prime}_{xx}-Q^{(1)}_{xx}&=&0\label{Q1xxeqn}\\
    \ell^2Q^{(1)\prime\prime}_{xy}-Q^{(1)}_{xy}+
    v_x^{(1)\prime}&=&0.\label{Q1xyeqn}
\end{eqnarray}
The Neumann boundary conditions on $Q_{ij}$ together with eqn~(\ref{Q1xxeqn}) imply that $Q^{(1)}_{xx}(y)=0$. Integrating eqn~(\ref{v1eqn}) yields $v_x^{(1)\prime}-\alpha_\mathrm{c}Q^{(1)}_{xy}=\sigma^{(1)}$, where  $\sigma^{(1)}$ is a constant. Eliminating  $v_x^{(1)}$ from eqn~(\ref{Q1xyeqn}) leads to 
\begin{equation}
    \ell^2Q_{xy}^{(1)\prime\prime}+(
    \alpha_\mathrm{c}-1)Q^{(1)}_{xy}=-
    \sigma^{(1)}.
\end{equation} 
To get a solution for $Q^{(1)}_{xy}$ that satisfies the Neumann boundary conditions, 
we must have 
\begin{eqnarray}
Q^{(1)}_{xy}&=&c_1\cos(\sqrt{
\alpha_\mathrm{c}-1}y/\ell)-\frac{
\sigma^{(1)}}{
\alpha_\mathrm{c}-1}\\
&=&c_1\cos\pi y-\frac{
\sigma^{(1)}}{
\alpha_\mathrm{c}-1}, 
\end{eqnarray}
Using eqn~(\ref{v1eqn}) and the no-slip boundary conditions implies $\sigma^{(1)}=0$ and $v^{(1)}=(c_1\alpha_\mathrm{c}/\pi)\sin\pi y$.
Note that to leading order, $v$ and $Q_{xy}$ are $\mathcal{O}(\delta\alpha^{1/2})$, but $Q_{xx}$ is at most $\mathcal{O}(\delta\alpha).$ At the next order, the equations are
\begin{eqnarray}
    v_x^{(2)\prime\prime}-(1+\pi^2\ell^2)Q_{xy}^{(2)\prime}&=&0\\
    -\ell^2Q_{xy}^{(2)\prime\prime}+Q_{xy}^{(2)}-
    v_x^{(2)\prime}&=&0\\
    -\ell^2Q_{xx}^{(2)\prime\prime}+Q_{xx}^{(2)}&=&c_1^2{(1+\pi^2\ell^2)}\cos^2\pi y,
\end{eqnarray}
with solutions 
\begin{eqnarray}
    Q_{xx}^{(2)}&=&c_1^2\frac{1+\pi^2\ell^2}{2
    }\left(1+
    \frac{\cos2\pi y}{1+4\pi^2\ell^2}\right)\\
    Q_{xy}^{(2)}&=&c_2\cos\pi y\\
    v_x^{(2)}&=&c_2\frac{1+\pi^2\ell^2}{
    \pi}\sin\pi y,
\end{eqnarray}
where $c_2$ is a constant.

To determine $c_1$, we must expand to $\mathcal{O}(\delta\alpha^{3/2})$:
\begin{eqnarray}    v_x^{(3)\prime\prime}-{(1+\pi^2\ell^2)}Q^{(3)\prime}_{xy}&=&-c_1\pi\sin(\pi y)\label{v3eqn}\\
    \ell^2Q_{xy}^{(3)\prime\prime}-Q_{xy}^{(3)}+
    v^{(3)\prime}&=&
    c_1^3C_0\left[\left(\frac{3}{2}+4\pi^2\ell^2\right)\cos\pi y\right.\nonumber\\
    &+&\left.\frac{1}{2}\cos3\pi y\right]\label{Qxy3},
\end{eqnarray}
where $C_0=(1+\pi^2\ell^2)^2/[2(1+4\pi^2\ell^2)
]$.
Integrating eqn~(\ref{v3eqn}) yields
\begin{equation}
    v_x^{(3)\prime}={(1+\pi^2\ell^2)}Q^{(3)}_{xy}
+c_1\cos\pi y+\sigma^{(3)},\label{stress3}\end{equation}
where the constant $\sigma^{(3)}$ appears in the expansion of the stress, $\sigma=v_x^\prime-\alpha Q_{xy}=\delta\alpha^{1/2}\sigma^{(1)}+\delta\alpha\sigma^{(2)}+\delta\alpha^{3/2}\sigma^{(3)}+\dots$. The solutions we have already found at lower order imply that $\sigma^{(1)}=\sigma^{(2)}=0$. The no-slip boundary conditions on $v_x^{(3)}$ also imply that $\sigma^{(3)}=0$. Thus, the stress vanishes not only at the critical value of the activity, but also as $\alpha$ is increased above $\alpha_\mathrm{c}$. Our numerical computations give the same result just above the critical activity. Using eqn~(\ref{stress3}) to eliminate $v^{(3)}$ from eqn~(\ref{Qxy3}) yields
\begin{equation}
\ell^2Q_{xy}^{(3)\prime\prime}+\pi^2\ell^2Q_{xy}^{(3)}=C_1\cos\pi y+C_2\cos3\pi y,\label{Qxy3eq}
\end{equation}
where
\begin{eqnarray}
    C_1&=&\frac{c_1^3(1+\pi^2\ell^2)^\trp{2}(3+8\pi^2\ell^2)}{4(1+4\pi^2\ell^2)
    }-c_1
    \\
    C_2&=&\frac{c_1^3(1+\pi^2\ell^2)^2}{4(1+4\pi^2\ell^2)
    }.
\end{eqnarray}

To find $c_1$, we use the Fredholm alternative,~\cite{StakgoldHolst2011} which implies that the right-hand side of eqn~(\ref{Qxy3eq}) must be orthogonal to the solution of the corresponding homogeneous equation. Thus, $C_1=0$, and
\begin{eqnarray}
Q_{xx}&=&\frac{2\delta\alpha}{\alpha_\mathrm{c}}\frac{
(1+4\pi^2\ell^2)}{3+8\pi^2\ell^2}\left(1+\frac{\cos2\pi y}{1+4\pi\ell^2}\right)+\mathcal{O}(\delta\alpha^{3/2})\\
Q_{xy}&=&\pm\frac{2\delta\alpha^{1/2}}{\alpha_\mathrm{c}}\sqrt{\frac{
(1+4\pi^2\ell^2)}{3+8\pi^2\ell^2}}\cos\pi y+\mathcal{O}(\delta\alpha)\\
v_x&=&\pm\frac{2\delta\alpha^{1/2}}{\pi}\sqrt{\frac{
(1+4\pi^2\ell^2)}{3+8\pi^2\ell^2}}\sin\pi y+\mathcal{O}(\delta\alpha),
\end{eqnarray}
where the two signs for $v_x$ and $Q_{xy}$ correspond to the two different spontaneous directions of flow, and the corresponding orientation of the directors. These analytical solutions agree well with our numerical solutions for the spontaneous unidirectional flow state with activity just above the critical activity. 

\section{Annular channel: nonlinear spontaneous flows}
\label{nonlinear annular}

\trp{In our work on the straight channel, we saw that simple shear flow led to a spatially uniform order parameter field $\mathsf{Q}$ when the activity is less than a critical value. Uniform  $\mathsf{Q}$ leads to zero active force on the fluid. In contrast, if the shear rate in the flow is spatially nonuniform, the alignment and degree of ordering of the directors will also be spatially nonuniform, leading to an active force. This situation arises in the case of curved boundaries---as in an annular channel---for any nonzero value of the activity, no matter how small.}
{Previous} \trp{theoretical} {studies involving curved boundaries}
\trp{have focused on the case of motionless walls. For example, Woodhouse and Goldstein found spontaneous circular flow in a circular chamber,\cite{woodhouse2012spontaneous}  and Norton et al. showed that the nature of topological defects in the director field is determined by the flow rather than the director anchoring conditions at the wall of a circular chamber.\cite{norton2018insensitivity}} 

\begin{figure*}[t]
\centering
\includegraphics[height=3.4in]{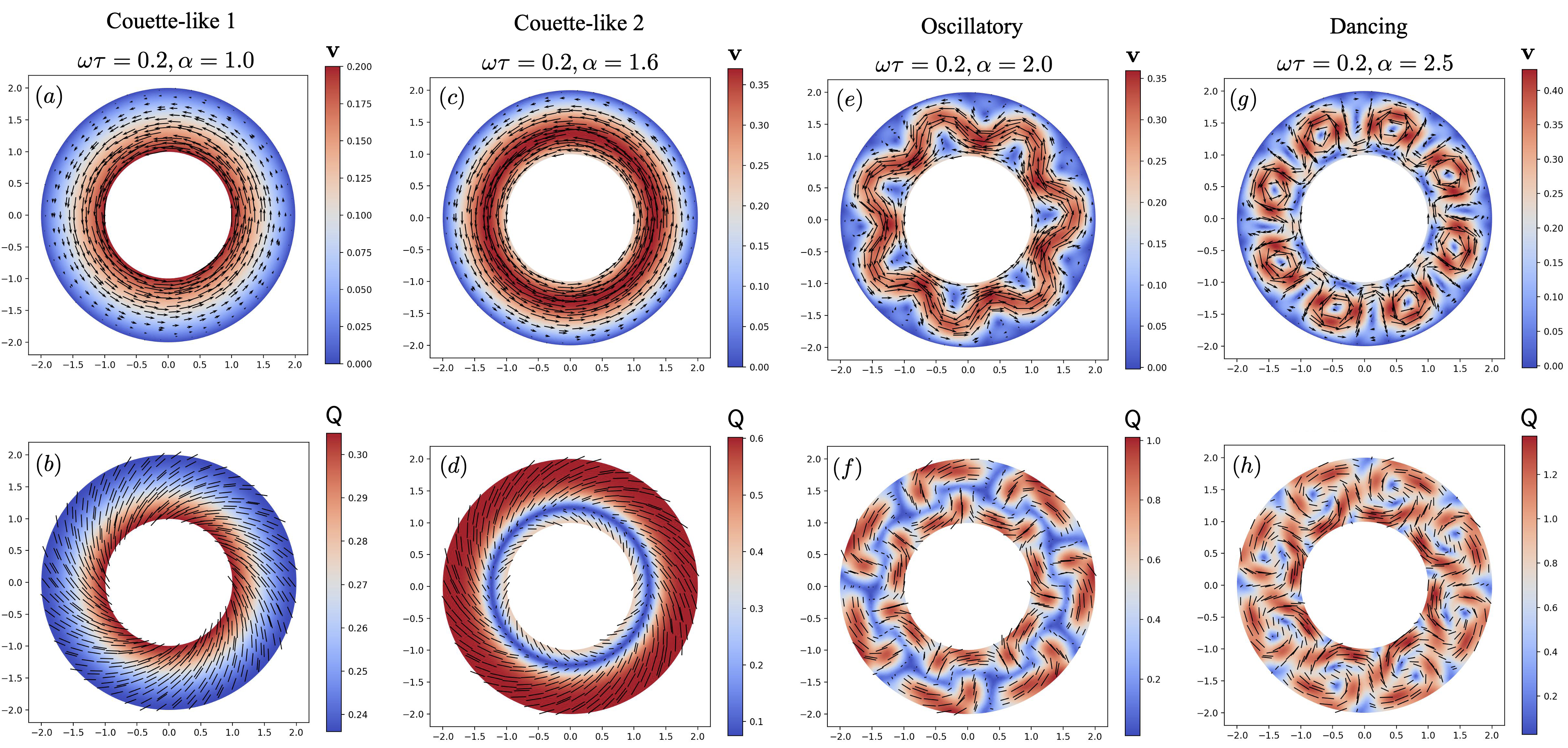}
\caption{
Examples of spontaneous flow states of an active gel in an annular channel. In these examples, the inner boundary rotates with dimensionless speed $\omega\tau=0.2$. The Couette-like 1 and 2 states are steady. In the oscillatory flow state, the flow pattern and order parameter configuration rotate at a steady rate.
The dancing flow state is 
unsteady. {Videos of the oscillatory and the dancing flow states can be found in the SI, sections S2.1 and S2.2 respectively}. 
In the velocity field plots (top panels),
colors denote flow speed 
and arrows denote 
flow direction. 
In the order parameter field plots 
(bottom panels), 
colors denote the scalar order parameter 
and lines denote the directors.
}
\label{fig:disk_states}
\end{figure*}
\trp{In this section,} we introduce curvature by considering 
the flow states of a two-dimensional active gel in 
\trp{the Taylor-Couette geometry of an} annular channel between two concentric circular boundaries of radius $R$ and $R+W$. We impose external shear 
by rotating the inner boundary with \trp{steady} angular frequency $\omega$, leaving the outer boundary stationary. 
Stokes flow in this geometry, \trp{known as} Couette flow, 
is  given by~\cite{landauFM}
\begin{equation}
    v_\theta = \frac{\omega R^2}{(2R+W)W}\left[\frac{(R+W)^2}{r}-r\right],\label{TCstokes}
\end{equation}
\trp{where $r$ is the radial polar coordinate. The second term of eqn~(\ref{TCstokes}) corresponds to rigid body rotation and does not lead to any strain rate, but the first term leads to a nonuniform strain rate, and thus induces a nonuniform order parameter field and an active force on the fluid for \textit{any} nonzero value of the activity.} To study the nonlinear flow states of active flows in the annular channel, we again employ the finite element software FEniCS to solve the the full nonlinear equations, eqns~(\ref{incompress})-(\ref{Qeqn}). 
We set $\ell=0.1$, $\lambda=1$ and $R/W=1$.

\subsection{Extensile fluids}
\begin{figure}[t]
\centering
\includegraphics[height=2.4in]{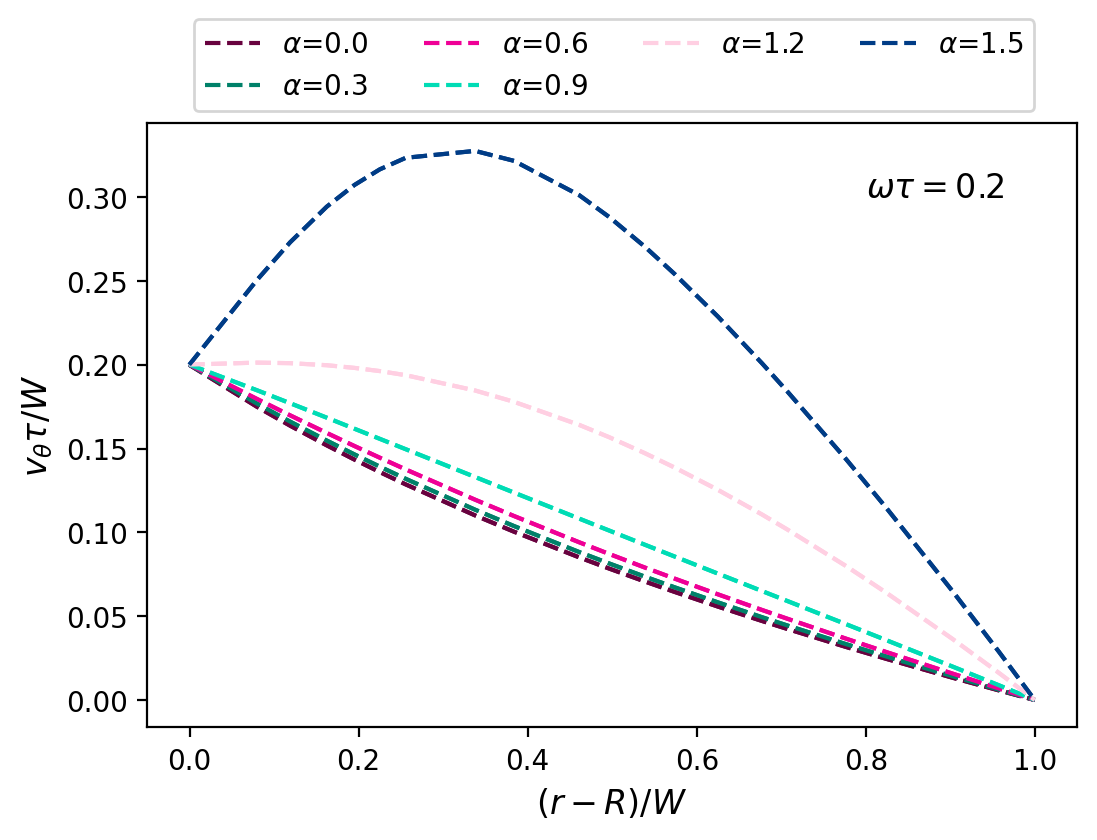}
\caption{  
Velocity profiles in dimensionless units for the extensile case of Couette ($\alpha=0$), Couette-like 1 ($\alpha=0.3$, $0.6$ and $0.9$) and Couette-like 2 ($\alpha=1.2$ and $1.5$) flow states, for $\omega\tau=0.2$.  
}
\label{fig:disk_extvx_a}
\end{figure}

We begin our discussion of the flow states in the annulus with extensile active gels, $\alpha>0$. As in the case of the straight channel, we give the initial director field some splay to induce counterclockwise or clockwise spontaneous flow, with the flow direction depending on the sense of the splay. For example, splay with the rods converging as we move counterclockwise around the annulus (Fig.~~\ref{fig:disk_states}d) leads to counterclockwise active flow (Fig.~\ref{fig:disk_states}c).
%
For the activities we used, we find the same kinds of active flow states as in the straight channel: Couette-like states which have no radial component of flow and are the analogs of the unidirectional states in the straight channel (Figs.~\ref{fig:disk_states}a--d), 
oscillatory states (Figs.~\ref{fig:disk_states}e and f), and dancing states (Figs.~\ref{fig:disk_states}g and h). 
We run the simulations until $t=600\tau$, and characterize the flow states as we did in the case of the straight channel (Sec.~\ref{sec:nonlinear}).
For the 
Couette-like flows, we distinguish 
two flow states by checking whether the maximum velocity is at the moving wall or in the interior of the annulus. If the flow is fastest on the wall, we label it a ``Couette-like 1'' flow state; otherwise the label is ``Couette-like 2''. If the transverse component of the velocity $v_r$ is nonnegligible, 
we check whether the torque exerted by the total flow on the inner boundary oscillates during the time interval $550\tau$-$600\tau$. If it oscillates, then the state is dancing, otherwise it is oscillatory. 
{There are a few flow states
near transitions 
that need a longer time to equilibrate. 
We also find multiple 
solutions for particular values of $\omega\tau$ and $\alpha$  for the oscillatory and dancing flows.
{Fig.~\ref{fig:disk_stability} shows flow transitions in the annular channel in range of $0 \leq \alpha \lesssim 2.5$ and $0 \leq \dot{\gamma}\tau \lesssim 1$.}
The transition from Couette-like to oscillatory flow states is relatively robust, with the transition states showing little dependence on the initial conditions. However, {comparing with} the case of the straight channel, the states observed in the transition from oscillatory to dancing flow are {more sensitive} to 
the choice of initial conditions.

\begin{figure}[t]
\centering
\includegraphics[height=2.6in]{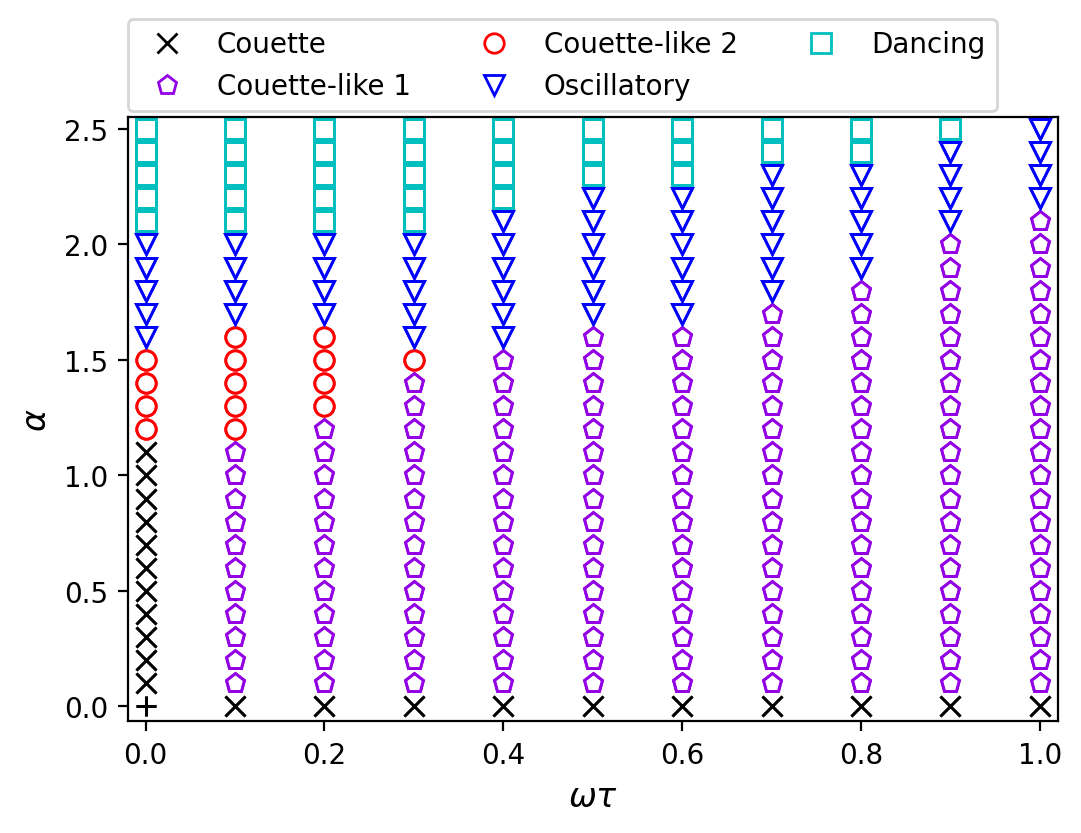}
\caption{  
Flow states for extensile fluids in an annular channel with  $\ell=0.1$ and $R/W=1$. The Couette states at $\omega=0$ are states of zero flow and zero order.
}
\label{fig:disk_stability}
\end{figure}

In the case of a straight channel, our numerical calculations always yielded the Newtonian simple shear state solution as long as the magnitude of the activity was small enough. The situation is different for the annular channel: our numerical calculations only yield the Newtonian Couette flow state solution (eqn (\ref{TCstokes})) when the activity vanishes. 
As emphasized earlier, any nonzero value of activity leads to active force and an active component of the flow because the order parameter field is nonuniform for nonzero wall rotation speed $\omega$. 
{
Green, Toner and Vitelli examined a similar phenomena for active nematics in which a surface of nonvanishing Gaussian curvature generates a spontaneous flow at arbitrarily low values of the activity parameter.\cite{green2017geometry}}
As long as $\omega\tau$ is sufficiently small, the flow profile varies continuously between the Couette, Couette-like 1, and Couette-like 2 states as the activity increases (Fig.~\ref{fig:disk_extvx_a}). Note that the flow velocity increases with activity for a given imposed rotation rate, as expected because extensile activity reduces the effective shear viscosity.\cite{hatwalne04}} Also, the change from the Newtonian Couette flow profile is small as long as the activity is modest, $\alpha\lesssim0.9$ (Fig.~\ref{fig:disk_extvx_a}), which we examine in Sec.~\ref{annular low shear}.

The oscillatory flow in an annular channel (Figs.~\ref{fig:disk_states}e and f; movies are in SI) is similar to the oscillatory flow in a straight channel. The flow and order parameter patterns are steady in  a frame that rotates at constant speed,
and the average volumetric flow rate ($\int\mathrm{d}rv_{\theta}/W$) is 
constant in time. Since we solve the equations in the annular domain without applying periodic boundary conditions, and still see steady rotation of the flow pattern and order parameter pattern, we can be confident that the constant wave speed we saw in the case of the oscillatory flows in the straight channel is  not
an artifact of the periodic boundary condition.

In the dancing state, the flow and order parameter patterns periodically change in 
time, 
similar to the case of the straight channel. 
Unlike the straight channel, the 
volumetric flow rate of the dancing flow state (Figs.~\ref{fig:disk_states}g and h; movies are in SI) in the annular case is 
not constant in time.
This time dependence arises because the difference in curvatures of the inner and outer boundaries of the annulus breaks the reflection symmetry of the boundaries of the straight channel that relates the dancing flow at the top wall to the dancing flow at the bottom wall.
Also, as in the straight channel, we observe 
moving pairs of $+1/2$ defect-like patterns \rap{with an} exchange \rap{of} partners in the annular dancing flow. {In the straight channel, the defect pairs are mirror images of each other (see Fig~\ref{fig:ext_dan}b), but in the annulus, the different curvatures of the two boundaries spoils this symmetry.}
\wl{Joshi et al. also found similar oscillatory and dancing flow states for active nematics by changing the curvatures of the annular channel without external shear.~\cite{joshi2023disks}} 

\begin{figure}[t]
\centering
\includegraphics[height=2.6in]{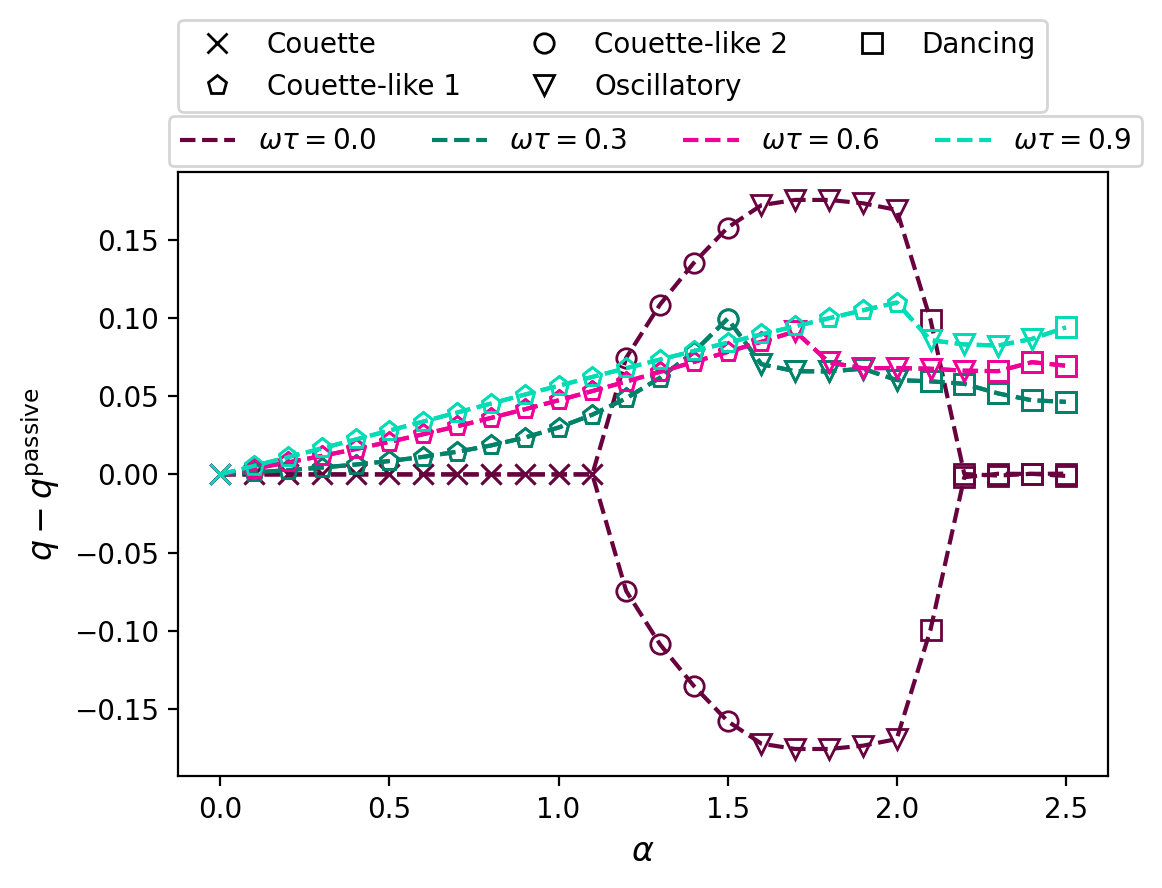}
\caption{  Active component of the volumetric flow rate for extensile fluids in dimensionless units of $v_{\theta} \tau/W$ as a function of activity in the annular channel. The symbols denote the flow states and the colors denote the externally imposed dimensionless shear rate. 
For the dancing flows, the square symbols show the average value of the oscillation of the 
volumetric flow rate.
}
\label{fig:disk_flux}
\end{figure}
Fig.~\ref{fig:disk_flux} shows the active component of the average flow rate (defined as before as the average flow rate of the total flow minus the average flow rate of the $a=0$ case) for the various flow states we studied in the annular channel. For the case of zero 
applied shear 
($\omega\tau=0$), there are 
positive and negative spontaneous flows when the activity 
exceeds a critical value. 
But for 
$\omega\tau\neq0$, 
the flow rate has no bifurcation: it continuously increases from zero as the activity increases from zero.
Another striking difference with the straight channel is that for nonzero rotation rates of the inner curved wall,
we only observe positive spontaneous flows 
(Fig.~\ref{fig:disk_flux}), 
even when we \trp{attempt to reverse the direction of flow by altering the initial conditions of the directors.}
This rectification arises 
because in the curved channel, the non-uniform alignment of the directors arising from the applied shear leads to spontaneous flow with the same rotation sense as the rotating wall.
Furthermore, since the wave translation direction corresponds to the direction of the spontaneous component of the flow, the oscillatory flow patterns all translate in the $+\theta$ direction when $\omega\tau\neq0$. 
{Another difference from the straight channel case is that the active contribution to the average flow rate does not disappear in the annular channel for larger shear rate.}

\begin{figure}[t]
\centering
\includegraphics[height=2.6in]{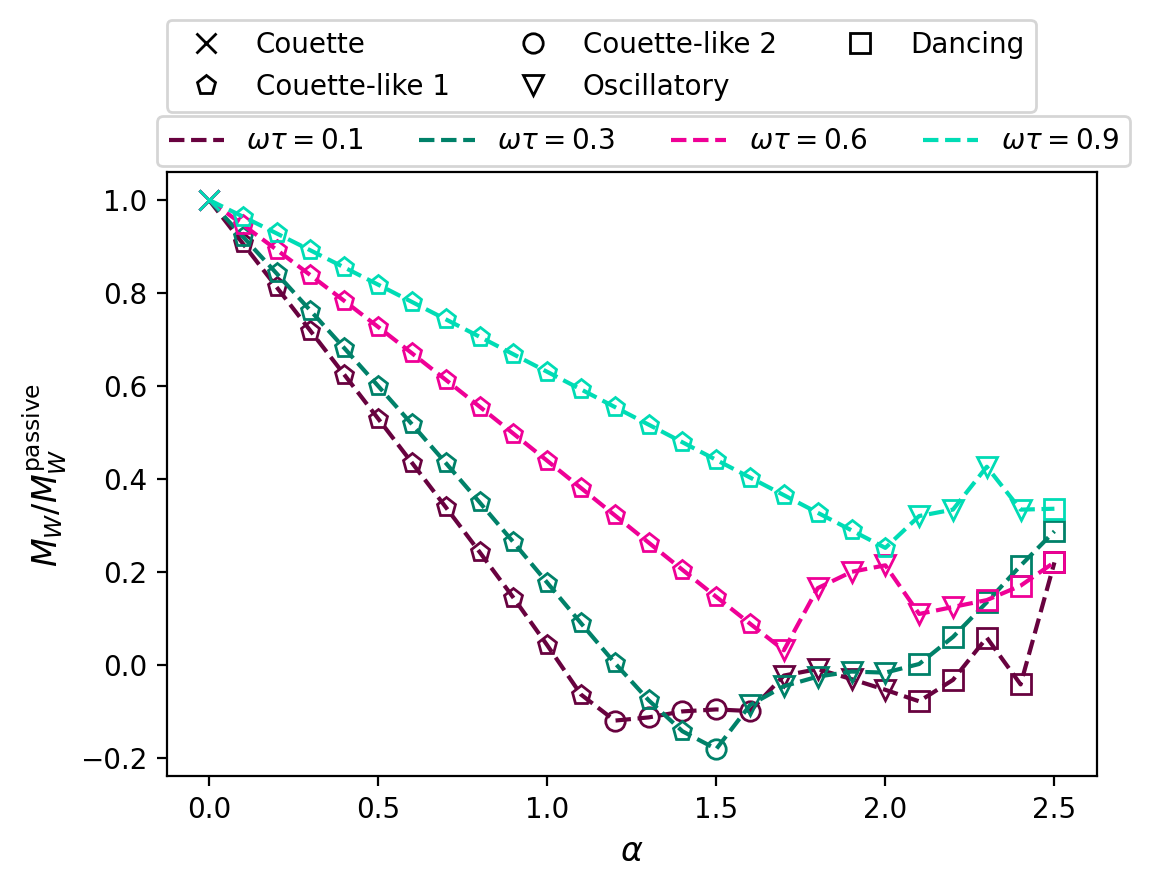}
\caption{
The torque imposed by the active flow of an extensile fluid on the
rotating disk normalized by the 
passive torque.
For the dancing flows, the square symbols 
denote the average value of the oscillation of the torque.
}
\label{fig:disk_torque}
\end{figure}

Fig.~\ref{fig:disk_torque} shows the torque exerted by the 
active fluid on the inner boundary, normalized by the wall torque in 
the passive case. The relation of the wall torque to the activity is very similar to the relation of the wall stress to the activity in the straight channel case, i.e. the normalized wall torque decreases with increasing activity for the Couette-like 1 flow state. 

The change in slope in the active-flow rate vs. $\alpha$ curve in Fig.~\ref{fig:disk_flux}, or the normalized wall torque vs. $\alpha$ in Fig.~\ref{fig:disk_torque} indicates the transition from the Couette-like flow state to the oscillatory flow state.
As noted earlier, sometimes our numerical approach finds oscillatory patterns of different wavelengths for the same values of the parameters\wl{, which would likely result in values of the volumetric flow rate and wall torque different from those shown in Fig.~\ref{fig:disk_flux} and Fig.~\ref{fig:disk_torque}.} Some of the variation in the normalized torque in the oscillatory and dancing flow regimes in Fig.~\ref{fig:disk_torque} arises from abrupt changes in wavelength as $\alpha$ was varied.

\begin{figure}[t]
\centering
\includegraphics[height=2.6in]{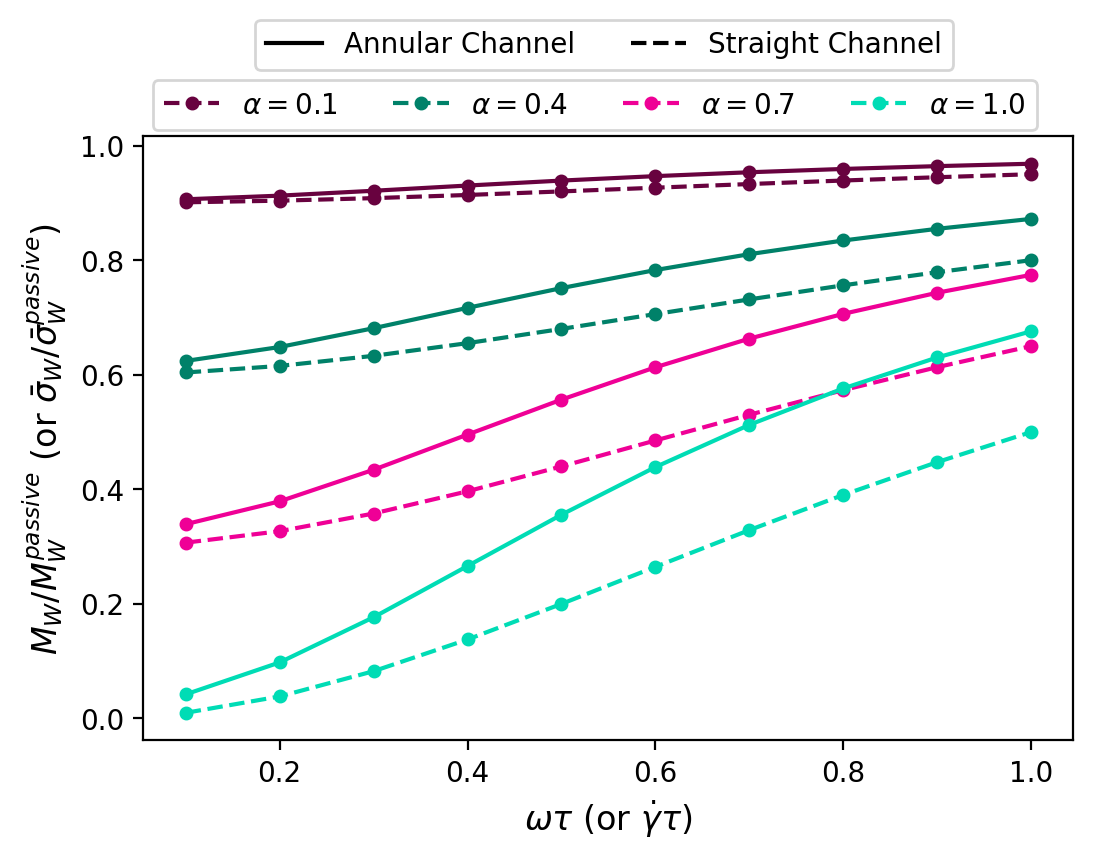}
\caption{  
Comparison of the normalized torque on the inner boundary of the annulus as a function of the dimensionless frequency of rotation of the inner disk and the normalized wall stress on the bottom wall of the straight channel as a function of the dimensionless shear rate. 
}
\label{fig:disk_torque_omega}
\end{figure}

We compare the wall torque and wall stress of annular and straight channels in Fig.~\ref{fig:disk_torque_omega} to show the effect of curvature on the wall stress as a function of external shear in the range of $0 < \alpha\leq 1$.
The normalized wall torque and wall stress are close to each other for small external shear rate and both increase with external shear rate, but the increase is larger in the annular channel, i.e. normalized wall torque is closer in value to the passive case. Thus, the curvature of the channel reduces the effect of activity on the wall with increasing external shear.


\begin{figure}[t]
\centering
\includegraphics[height=2.35in]{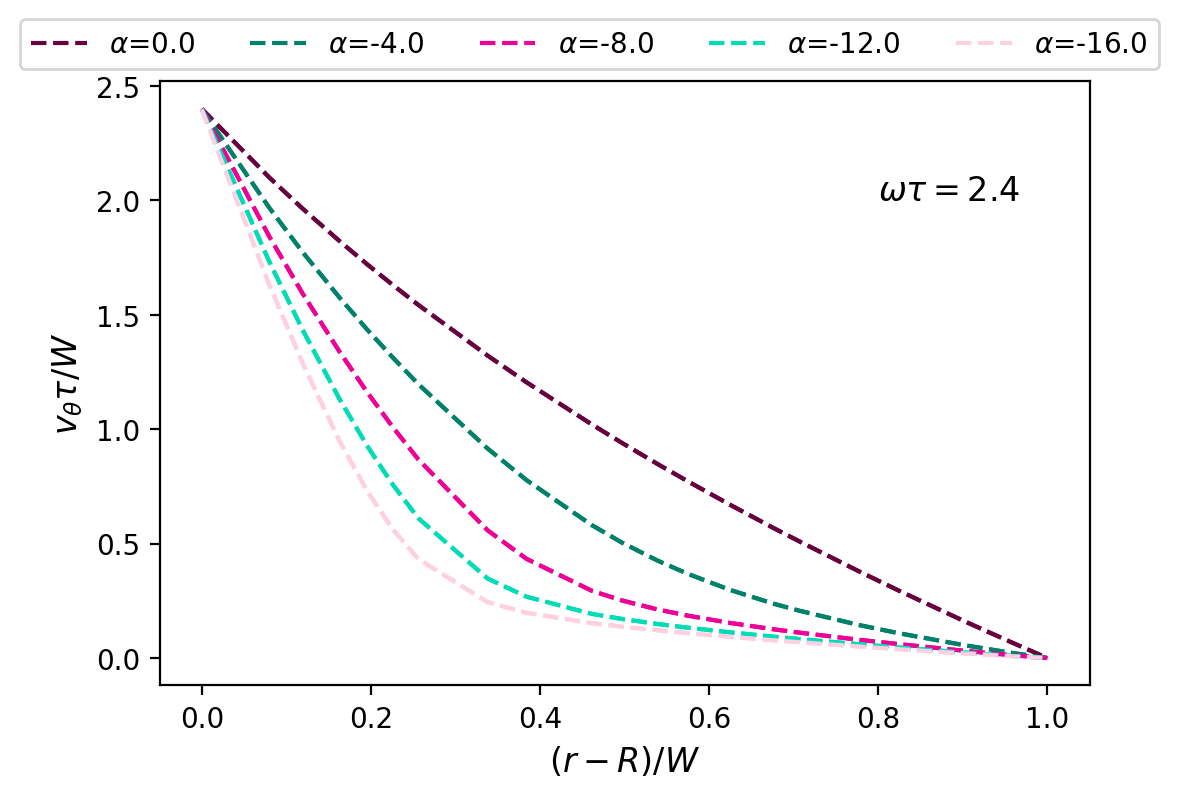}
\caption{  
Velocity profiles in dimensionless units for  Couette flow  ($\alpha=0$) and the contractile Couette-like 1 ($\alpha<0$) flows states, 
for $\omega\tau=2.4$. 
}
\label{fig:disk_convx_a}
\end{figure}

\subsection{Contractile fluids}

We studied contractile active fluids 
in a two-dimensional annulus with the parameters in the range $-16 \leq \alpha < 0$ and $0 < \omega \tau \leq 2.4$. When $\alpha<0$, we only found Couette-like states with no radial component of the flow. Since contractile activity is effectively shear thickening,\cite{hatwalne04}  the effect of the activity is always to reduce the flow relative to Newtonian Couette flow (Fig.~\ref{fig:disk_convx_a}). As in the extensile case, the direction that the active component of the flow travels around the annulus is independent of the initial conditions, but unlike the extensile case, the active component of flow is negative (against the direction imposed by the externally applied shear). The magnitude of the negative flow is always less than the magnitude of the externally imposed Couette flow; therefore, the total flow never reverses. In this sense, the contractile annular flows are Couette-like 1 states rather than Couette-like 2 states.
In accord with the effective shear-thickenening  of contractile active fluids, the total torque on the inner boundary is always greater than the hydrodynamic torque in  Couette flow (SI Fig.~S4).

\begin{figure}[t]
\centering
\includegraphics[height=4.3in]{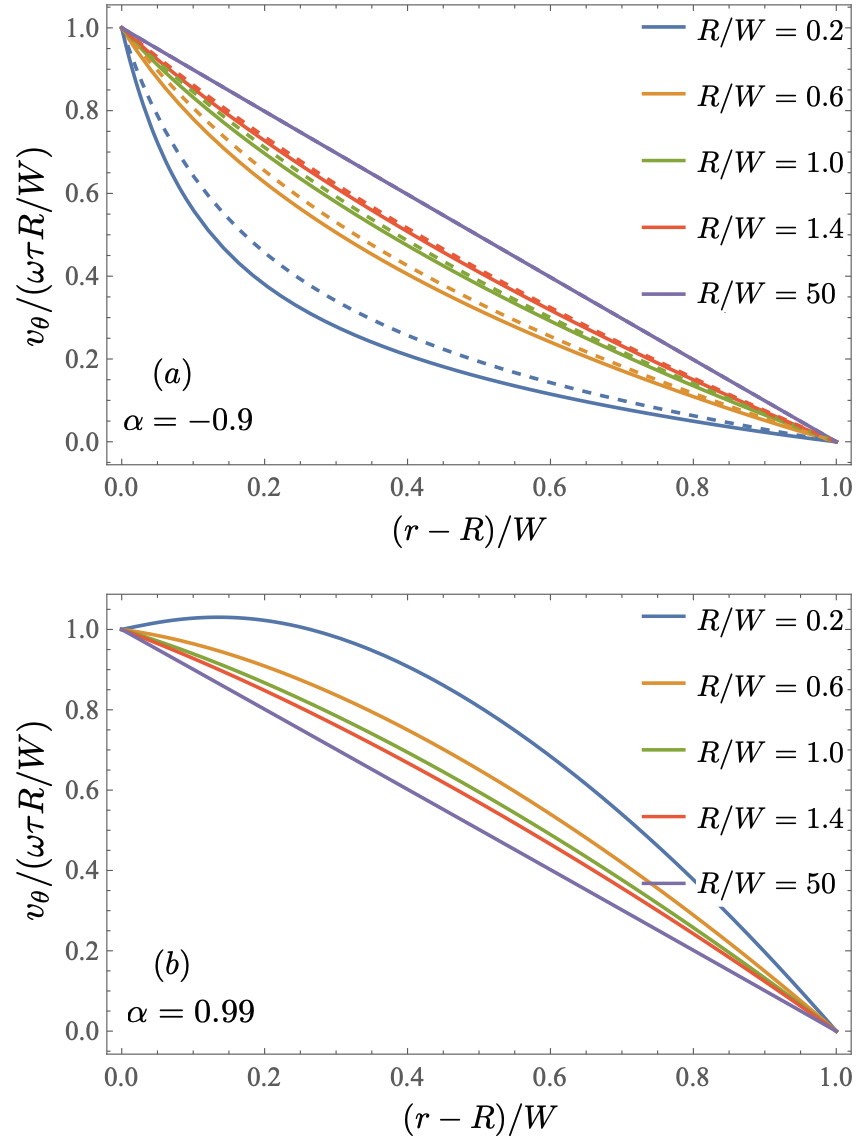}
\caption{Analytical results for flow velocity of the Couette-like 1 state in an annular channel with weak order for $\ell=0.1$, $\lambda=1$, and various values of $R/W$ for (a) a contractile fluid with $\alpha =-0.9$ and (b) an extensile fluid with $\alpha=0.99$. The dashed lines in the panel (a) show the results for Newtonian flows ($\alpha=0$) for comparison.
}
\label{fig:linear_disk_R}
\end{figure}


\begin{figure}[t]
\centering
\includegraphics[height=2.2in]{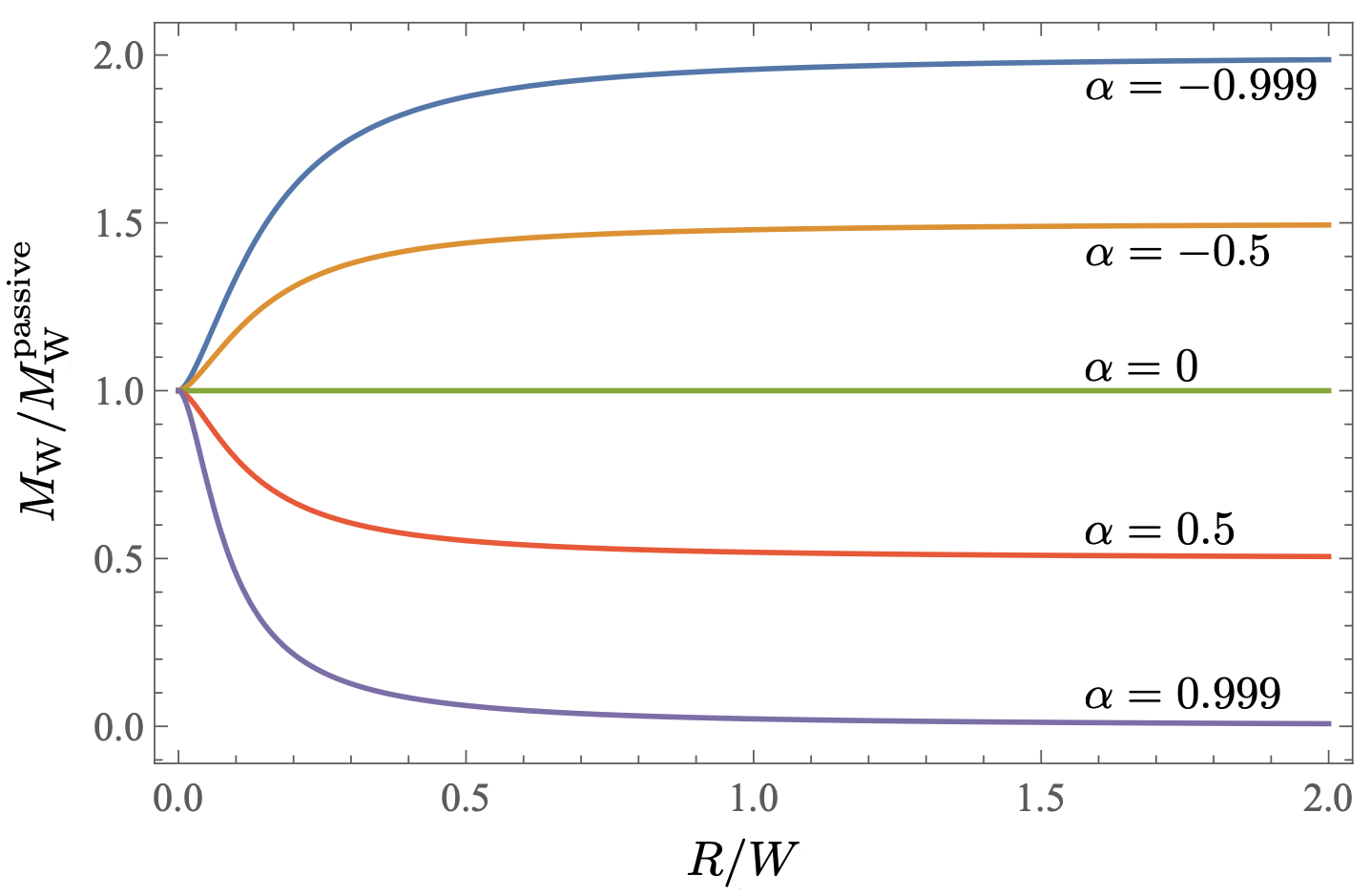}
\caption{Dependence of normalized wall torque on $R/W$ for different $\alpha$ from linear analysis at low shear with $\ell=0.1$ and $\lambda=1$. The colors denote different activities.
}
\label{fig:linear_disk_M}
\end{figure}

\section{Annular channel: Linear analysis of curvature at low shear rate}\label{annular low shear}

In this section we study the limit in which the flow in the annular channel is slow enough that the induced order is small, $S\ll1$. For slow enough flow, it is valid to neglect the nonlinear terms in eqn (\ref{Qeqn}).
On the one hand, 
this analysis 
offers a theoretical explanation of some of the observations in Sec.~\ref{nonlinear annular}; on the other hand, it gives some insight into the role of the curvature of the boundaries, which we did not vary in the previous section. 
For convenience, here we restate the modified Stokes equation (eqn~(\ref{veqn})) in dimensionless form, along with the dimensionless form of the steady linearized equation for $\mathsf{Q}$:
\begin{eqnarray}
0&=&-\boldsymbol{\nabla}p+\nabla^2\mathbf{v}-{\frac{\alpha}{\lambda}}\nabla\cdot\textsf{Q}\label{v2eqn}\\
0&=&-\textsf{Q}+\ell^2\nabla^2\mathsf{Q} 
+2 \lambda \mathsf{E},\label{Q2eqn}
\end{eqnarray}
As in our numerical calculations, we use the width $W$ of the channel as the unit of length.
Since we seek to study the Couette-like flow state, we assume $\mathbf{v}=v_\theta(r)\hat{\bm\theta}$. Note that this flow is incompressible. We also suppose that $p$, $Q_{rr}$, and $Q_{r\theta}$ are functions of radius only. With these assumptions, the $rr$ component of eqn~(\ref{Q2eqn}) is homogeneous, which together with the Neumann boundary conditions $\partial_rQ_{rr}=0$ at $r=R/W$ and $r=R/W+1$ implies $Q_{rr}=0$. Since the radial component of the modified Stokes equation(eqn~(\ref{v2eqn}) with $Q_{rr}=0$ implies that the pressure gradient vanishes, we take $p=0$.

To solve for the velocity and order parameter fields, we take the divergence of eqn~(\ref{Q2eqn}) and combine with eqn (\ref{v2eqn}) with $p=0$ to find
\begin{equation}
    \nabla^2\left(\nabla^2-\frac{1}{\xi^2}\right)\mathbf{v}=0,
\end{equation}
where $\xi^2={\ell^2}/(1-\alpha)$. To focus our attention on the Couette-like states only, we restrict our analysis to $\alpha<1$ in this section. Thus, 
\begin{eqnarray}
    v_\theta=c_1 r+c_2/r+c_3 I_1(r/\xi)+c_4 K_1(r/\xi),\label{vc1234}
\end{eqnarray}
where the $c_i$ are constants to be determined, and $I_1(x)$ and $K_1(x)$ are modified Bessel functions. Inserting the velocity field eqn (\ref{vc1234}) into the $r\theta$ component of eqn~(\ref{Q2eqn}),
\begin{equation}
    0=\ell^2 \left(Q_{r\theta}''+\frac{1}{r}Q_{r\theta}'-\frac{4}{r^2}Q_{r\theta}\right)-Q_{r\theta}+\lambda\left(v_\theta^\prime-\frac{v_\theta}{r}\right),\label{Qrthetav}
\end{equation}
yields
\begin{eqnarray}
    &&\ell^2\left(Q_{r\theta}''+\frac{1}{r}Q_{r\theta}'-\frac{4}{r^2}Q_{r\theta}\right)-Q_{r\theta}
    \nonumber\\
    &=&\frac{\lambda}{\xi}
    \left[\frac{2c_2\xi}{r^2}-c_3 I_2(r/\xi)+c_4 K_2(r/\xi)\right],
    \label{Qrthetav2}
\end{eqnarray}
which has general solution
\begin{eqnarray}
    Q_{r\theta}&=&c_5 I_2(r/\ell)+c_6 K_2(r/\ell)-2c_2\lambda/r^2\nonumber\\
    &-&c_3\frac{\lambda\xi}{\ell^2-\xi^2} I_2(r/\xi)+c_4\frac{\lambda \xi}{\ell^2-\xi^2}K_2(r/\xi).
\end{eqnarray}
Inserting this solution into the modified Stokes equation~[eqn~(\ref{v2eqn})] shows that $c_5=c_6=0$. The rest of the integration constants are determined by the no-slip boundary conditions on the (dimensionless) velocity, $v_\theta(R/W)=\omega \tau R/W$ and $v_\theta(R/W+1)=0$, and the Neumann boundary conditions on the order parameter field $Q_{rr}^\prime(R/W)=Q_{rr}^\prime(R/W+1)=0$.
The 
complete formulas are too 
complicated to display, but we plot the velocity in Fig.~\ref{fig:linear_disk_R} for various ratios of $R/W$ for a representative contractile case (top panel) and extensile case (bottom panel). In both cases, the flow velocity approaches a linear profile as $R/W$ becomes large, as expected, since in that limit the curvature of the annulus becomes unimportant, and the flow approaches simple shear flow. For the contractile case, Fig.~\ref{fig:linear_disk_R}a, the velocity profile is close to the Newtonian result, with the agreement between the two cases getting better as $R/W$ increases. For the extensile case, the velocity curves for different values of $R/W$ get closer to each other as $\alpha$ increases, becoming very close to the linear profile around $\alpha=0.885$. Above this value of activity, the order of the curves reverses, with the linear curve lying below all the other curves. When $\alpha$ gets very close to unity and $R/W$ is small, the maximum velocity is not at the wall, i.e. the flow continuously changes from the Couette-like 1 state to the Couette-like 2 state [Fig.~\ref{fig:linear_disk_R}b]. Fig.~\ref{fig:linear_disk_M} shows the total torque $M=2\pi R^2\sigma_{r\theta}$ on the circle $r=R$ as a function of $R/W$. Note that the limit of a straight channel is almost obtained when $R$ becomes comparable to $W$. The torque for a contractile fluid is higher than the passive value since contractile fluids effectively increase the shear viscosity. Likewise, the torque for an extensile fluid is less than the passive value since extensile fluids are shear thinning. The torque approaches the passive value when $R\ll W$. Note that since we use $W$ as the unit of length, the limit $R\ll W$ corresponds to making the inner cylinder of vanishing thickness. When $R<\ell$, the term $\ell^2\nabla^2\mathsf{Q}$ dominates eqn~(\ref{Q2eqn}), and therefore $\mathsf{Q}\rightarrow0$. In this limit, the active force vanishes, and flow is Couette flow.

It is informative to find the velocity and the order parameter field in the limit $R\gg W$, where the curvature of the annulus is small. Rather than taking the limit of the formulas used to make Figs.~\ref{fig:linear_disk_R} and \ref{fig:linear_disk_M}, it is simplest to solve the equations directly using regular perturbation theory in powers of $W/R$. 
Reinstating the dimensions and writing $r=R+y$, we find
\begin{eqnarray}
    v_\theta&=&\omega R\left(1-\frac{y}{W}\right) -\frac{\omega y}{2}\left(1-\frac{y}{W}\right)\nonumber\\
    &+&\frac{2\alpha\ell^2\omega W}{1-\alpha}\left[\frac{1-\cosh\left[{(1-2y/W)/}{\xi}\right]}{\cosh[{1}/(2\xi)]}\right]\label{vthetaasympt}\\
    Q_{r\theta}&=&-\frac{\lambda \omega \tau R}{W}\nonumber\\
    &+&\frac{\lambda\omega\tau}{2}\left[4y/W-3+4\xi\frac{\sinh\left[{(1-2y/W)/}{\xi}\right]}{\cosh[{1}/(2\xi)]}\right].\label{Qrthetaasympt}
\end{eqnarray}
The first terms of eqns~(\ref{vthetaasympt}) and (\ref{Qrthetaasympt}) correspond to the velocity and order parameter field, respectively, of a straight channel with an infinitesimal imposed shear rate $\dot{\gamma}=\omega R/W$. The remaining terms are the corrections due to the nonzero curvature of the annular channel. Unlike our weakly analysis of the active flow in the straight channel (Sec.~\ref{weaknonlinear}), which had spontaneous flow in either direction, here we see that the component of flow driven by the activity has a definite sign, and is the same direction as externally imposed flow for extensile fluids.

\section{Summary}
We investigated the stability and flow states of the active gel confined in a channel subject to a external shear.
An externally imposed shear flow can stabilize an extensile fluid that would be unstable to spontaneous flow when there is no external shear flow, and destabilize a contractile fluid that would be stable against spontaneous flow when there is no external shear flow. In accordance with previous simulations \cite{samui2021flow,VargheseBaskaranHaganBaskaran2020} \rap{carried out in the absence of external shear,} we find three kinds of nonlinear flow states in the range of parameters we study: unidirectional flows, oscillatory flows, and dancing flows for extensile fluids. The unidirectional flow observed in the straight channel can have a spontaneous active component which is either positive---in the same direction as the moving wall---or negative---in the opposite direction of the moving wall.
The oscillatory flow states also have two possible directions for the spontaneous active component when the externally imposed shear rate is small. For greater imposed shear rates, the spontaneous flow direction will be the same as the moving wall. For contractile gels, we only observe 
unidirectional flow states in the range of parameters that we studied. These unidirectional flows can have positive or negative spontaneous active components.
In the analysis of the the wall stress caused by the active flow on the moving boundary, the extensile flow helps the motion of the moving boundary, while the contractile flow resists the motion. Moreover, the external shear flow can weaken this effect of activity on the motion.

Our analysis of the curvature shows there are three main differences between
the flows states for the straight channel and the annular channel. First 
in the annular channel, there is no critical activity for the system to be stable against the spontaneous flow given a nonzero external shear. Second, we only observe one direction of spontaneous flow: positive for extensile gels, but negative for contractile gels.
Last, the average volumetric flow rate of the annular case oscillates with time for the dancing flow state, while it is steady 
in the straight channel.  
Also, we find increasing the curvature of the streamlines weakens the dependence of the wall stress on activity.

Our work suggests several directions for future study. An obvious extension is to work in three dimensions, allowing both the directors and velocity vectors to point out of the plane and vary in both directions across a channel. Also, it would be natural to study the effect of aligning flows induced by a pressure gradient rather than a moving wall, since Poiseuille-like flow may be easier to study experimentally.

\section*{Acknowledgements} This work was supported in part by the National Science Foundation through Grant Nos. MRSEC DMR-2011846, CBET-2227361, and  PHY-1748958. We are grateful to Jesse Ault, Kenny Breuer, Guillaume Duclos, Hamid Karani, Jasper Chen, Alexander Morozov, and Pranay Sampat for helpful discussions. We also thank the Center for Computation and Visualization (CCV) at Brown university for use of high performance computing facilities.

\bibliography{newrefs}
\bibliographystyle{rsc}

\end{document}


\maketitle
\begin{center}
\noindent\large{Wan Luo$^{\ast}$\textit{$^{ab}$}, Aparna Baskaran\textit{$^{c}$}, Robert A. Pelcovits\textit{$^{de}$}, and Thomas R. Powers\textit{$^{abde\dag}$}}
\end{center}
\vspace{2em}
\textit{$^{\ast}$~Email: Wan\_Luo@brown.edu}\\
\textit{$^{\dag}$~Email: Thomas\_Powers@brown.edu}\\
$^{a}$~School of Engineering, Brown University, Providence, RI 02912, USA.\\
$^{b}$~Center for Fluid Mechanics, Brown University,Providence, RI 02912, USA.\\ 
$^{c}$~Martin Fisher School of Physics, Brandeis University, Waltham, MA 02453 USA.\\
$^{d}$~Department of Physics, Brown University, Providence, RI 02912, USA.\\
$^{e}$~Brown Theoretical Physics Center, Brown University, Providence, RI 02912, USA.
\\
\\
\\

Movies of unsteady flow states in both the straight and annular channel cases are presented in this supplementary information. Also, more details about the contractile flows are shown, including the average flow rate, the wall stress in the straight channel case and the wall torque in the annular channel case. The supplementary information is divided into two sections: \ref{straight} Straight channel and \ref{annular} Annular channel.

\newpage
\section{Straight channel}
\label{straight}
\subsection{Oscillatory flow state}
A movie of the velocity field for positive spontaneous flow (Fig. 9a) is shown in the Movie S1: stra-posi-osci-U.gif.
Note that in the title of the movie 
$t$ is time in units of $\tau$, $a$ is activity in units of $\eta/(\tau\lambda)$, and $u0$ is moving velocity of the bottom plate of the straight channel (i.e. $\dot{\gamma} W$ in the main text) in units of $W/\tau$. The director and scalar order parameter fields for positive spontaneous flow (Fig. 9b) are shown in the Movie S2: stra-posi-osci-Q.gif. 

The velocity field for negative spontaneous flow (Fig. 9c) is shown in the Movie S3: stra-negt-osci-U.gif.
The director and scalar order parameter fields for negative spontaneous flow (Fig. 9d) are shown in the Movie S4: stra-negt-osci-Q.gif.

\subsection{Oscillatory-like flow state}
An example of a dynamical final state for oscillatory-like flow with $\dot{\gamma}=0$ and $\alpha=2.3$. 
The flow pattern is not perfectly periodic in the horizontal direction. 
A movie of the velocity field is shown in the Movie S5: stra-oscilike-U.gif.
A movie of the  director and scalar order parameter fields is shown in the Movie S6: stra-oscilike-Q.gif.
\subsection{Dancing flow state}
Movies for the example of the dancing flow state shown in Fig. 11. A movie of the velocity field is shown in the Movie S7: stra-danc-U.gif.
A movie of the director and scalar order parameter fields is shown in the Movie S8: stra-danc-Q.gif.
\subsection{Flow rate and wall stress of contractile fluids}
The steady-state of activity-driven volumetric flow rate and average wall stress imposed by the active flow on the bottom plates of the channel for contractile fluids.
\begin{figure}[H]
\centering
\includegraphics[height=3.2in]{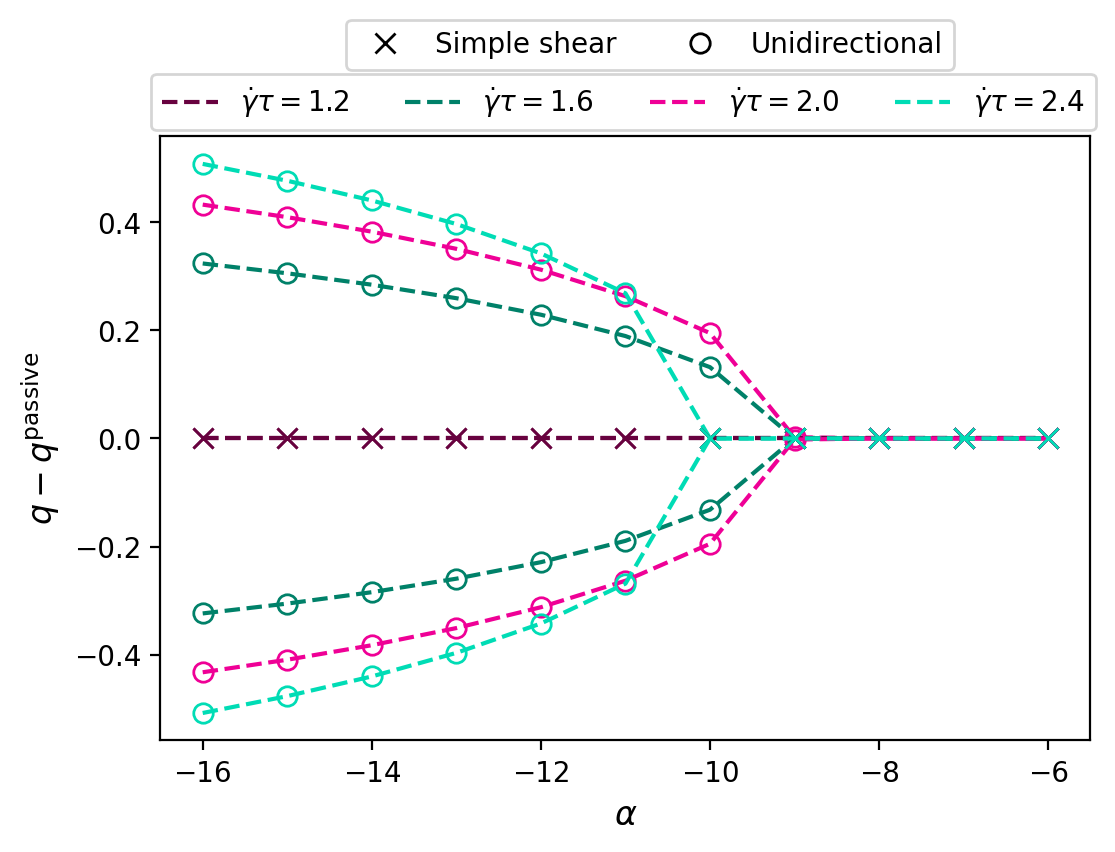}
\caption{
Activity-driven volumetric flow rate
in contractile fluids in the straight channel. Symbols denote the flow state and colors denote the external shear rates. The plot shows results for the two kinds of initial conditions of the director field with a positive and negative $x$ component.}
\label{fig:cont_flux}
\end{figure}

\begin{figure}[H]
\centering
\includegraphics[width=4.1in]{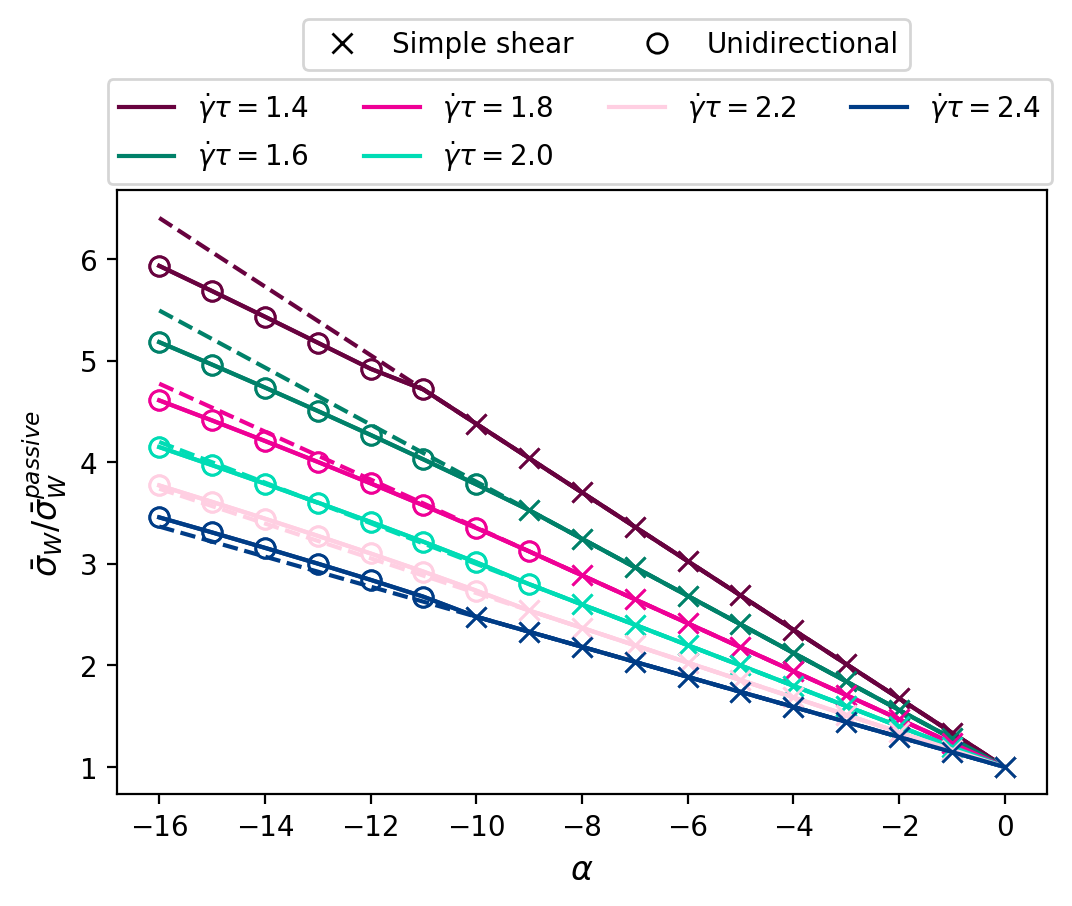}
\caption{
The average wall stress imposed by the active flow on the bottom plates of the straight channel $\bar{\sigma}_W$ as a function of dimensionless activity $\alpha$ for contractile fluids. 
The dashed lines are extended lines of simple shear flow. They are plotted for comparing the different slopes of simple shear flow and unidirectional flows. The results in the figure are insensitive to the two kinds of initial conditions for the director field, i.e., $\pm x$ splay-like alignment.
}
\label{fig:cont_stress}
\end{figure}

\newpage
\section{Annular channel}
\label{annular}
\subsection{Oscillatory flow state}
A movie of the velocity field for positive spontaneous flow (Fig. 16e) is shown in the Movie S9: annu-osci-U.gif.
Note that in the title of the movie 
$t$ is time in units of $\tau$, $a$ is activity in units of $\eta/(\tau\lambda)$, and $\omega$ is rotation frequency of the inner boundary of the annular channel in units of $1/\tau$. The director and scalar order parameter fields for positive spontaneous flow (Fig. 16f) are shown in the Movie S10: annu-osci-Q.gif.

\subsection{Dancing flow state}
A movie of the velocity field of the dancing flow state (Fig. 16g) is shown in the Movie S11: annu-danc-U.gif.
Director and scalar order parameter fields of the dancing flow  state (Fig. 16h) are shown in the Movie S12: annu-danc-Q.gif.

\subsection{Flow rate and wall torque of contractile fluids}
Steady-state of activity-driven volumetric flow rate and wall torque imposed by the active flow on the inner disk of the annular channel in contractile fluids.
\begin{figure}[h]
\centering
\includegraphics[height=3.4in]{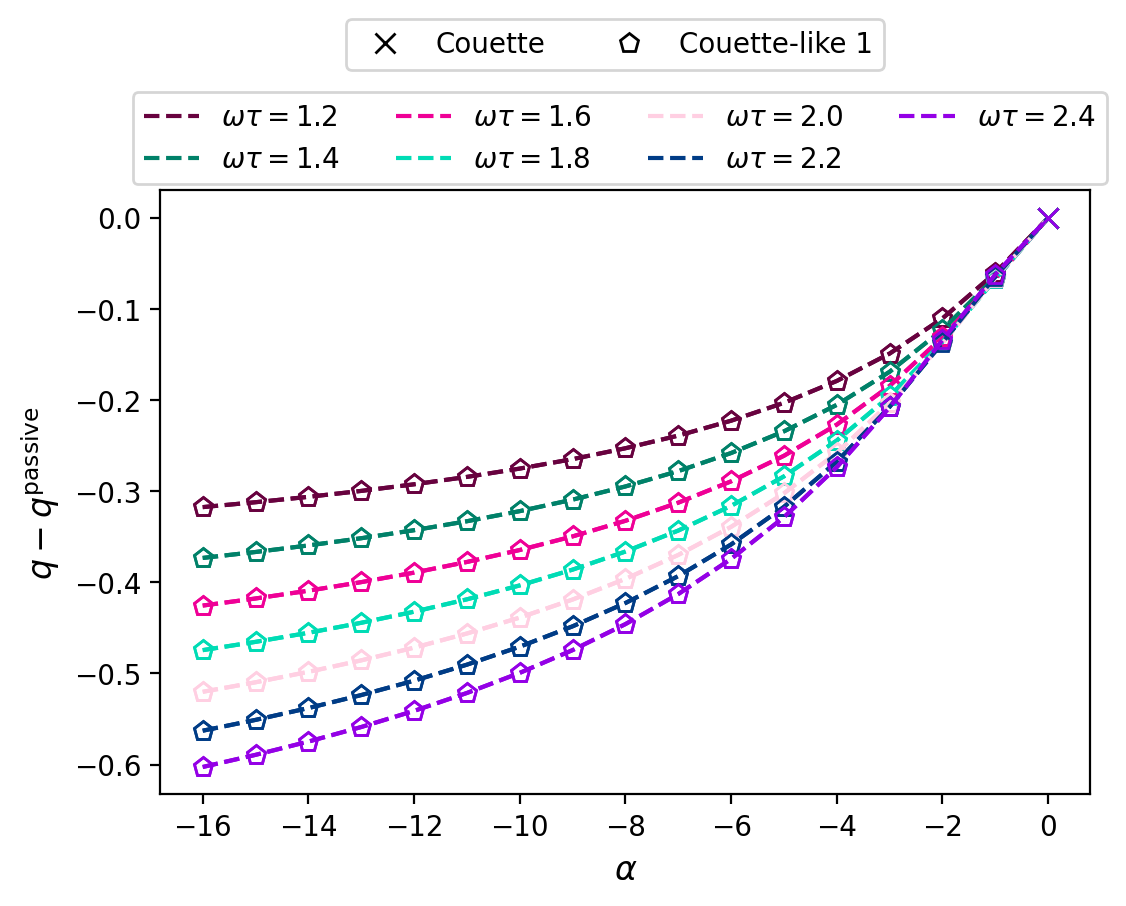}
\caption{Activity-induced flux for contractile fluids in dimensionless units of $v_{\theta} \tau/W$  as a function of activity in the annular channel. The plot shows results for the two kinds of initial conditions of the director field with a positive and negative $\theta$ component.
}
\label{fig:disk_cont_flux}
\end{figure}

\begin{figure}[h]
\centering
\includegraphics[height=3.4in]{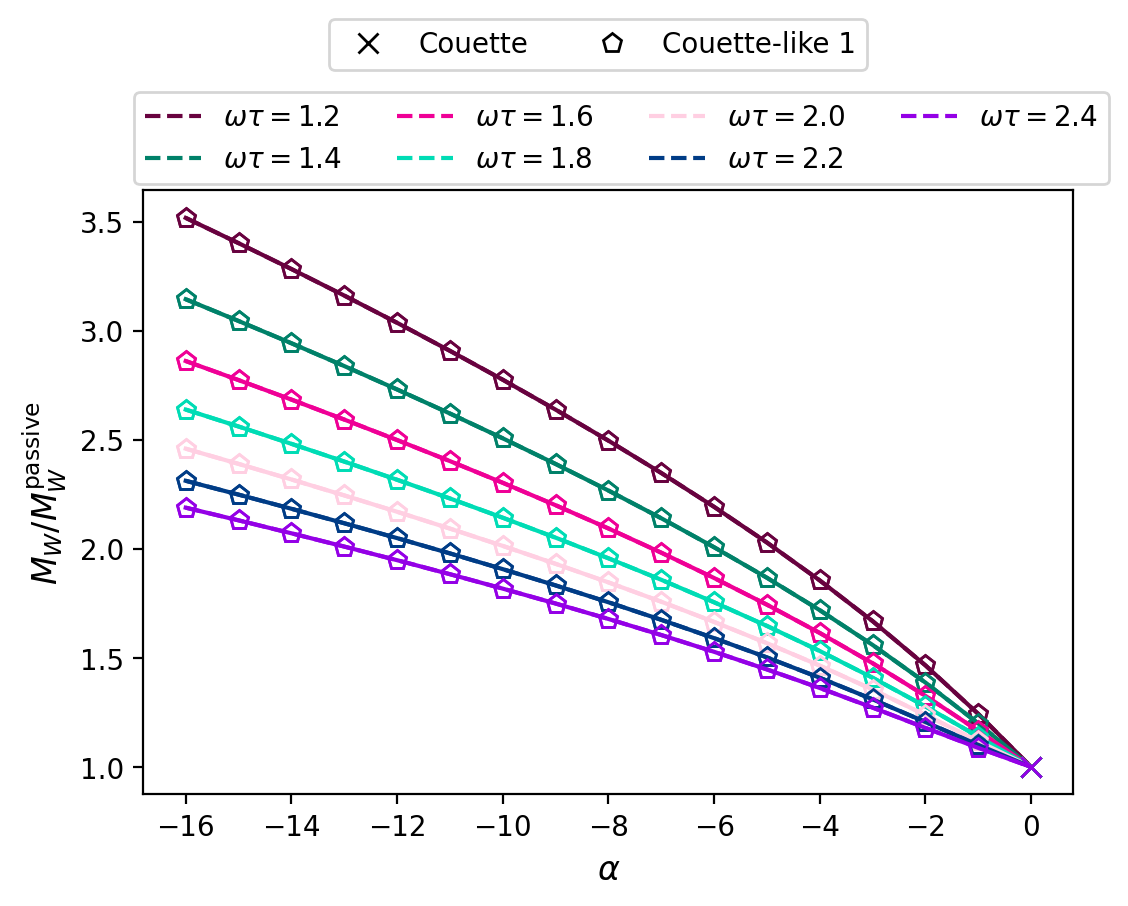}
\caption{ 
The wall torque of different active flows normalized by the passive flow 
in contractile fluids in the annular channel. The results in the figure are insensitive to the two kinds of initial conditions for the director field, i.e., clockwise and counterclockwise splay-like alignment.
}
\label{fig:disk_cont_stress}
\end{figure}